\documentstyle[amsmath,graphics,graphicx,rotate,pifont,psfig]{l-aa}
\input{epsf}

\begin{document}
   \thesaurus{08         
              (02.08.1; 
               02.19.1; 
               02.20.1;
               09.03.1;
               09.11.1 )}

\title{The shock waves in decaying supersonic turbulence}

\author{Michael D. Smith$^{1}$, Mordecai-Mark Mac Low$^{2,3}$ \& 
Julia M. Zuev$^4$}
\offprints{M.D. Smith}

\institute{
$^1$ Armagh Observatory, College Hill, Armagh BT61 9DG, Northern Ireland\\
$^2$ Max-Planck-Institut f\"ur Astronomie, K\"onigstuhl 17,
D-69117 Heidelberg, Germany\\ 
$^3$ Department of Astrophysics, American Museum of Natural History,
79th St. at Central Park West, New York, New York, 10024-5192, USA\\
$^4$ JILA, University of Colorado, Boulder, Campus Box 440, Boulder, CO 80309, USA\\ 
Internet: mds@star.arm.ac.uk, mordecai@amnh.org, julia.zuev@colorado.edu}

\date{Received; accepted}

\maketitle
\markboth{Smith, Mac Low \& Zuev: Supersonic Turbulence}{}

\begin{abstract} 

We here analyse numerical simulations of supersonic, hypersonic 
and magnetohydrodynamic turbulence that is free to decay. Our 
goals are to understand the dynamics of the decay and the
characteristic properties of the shock waves produced.  This 
will be useful for interpretation of  observations of both  
motions in molecular clouds and sources of non-thermal radiation.  

We find that  decaying hypersonic turbulence possesses an 
exponential tail of fast shocks and an exponential decay in time, 
i.e. the number of shocks is proportional to $t\,\exp (-ktv)$ for 
shock velocity jump $v$ and mean initial wavenumber $k$. In 
contrast to the velocity gradients, the  velocity Probability 
Distribution Function remains Gaussian with a more complex decay 
law. 

The energy is dissipated not by fast shocks but  by a large number of 
low Mach number shocks. The power loss peaks near a low-speed
turn-over in an exponential distribution. An analytical extension of 
the mapping closure technique  is able to predict the basic decay 
features. Our analytic description of the distribution of shock 
strengths should prove useful for direct modeling of observable 
emission. We note that an exponential distribution of shocks such as 
we find will, in general, generate very low excitation  shock 
signatures. 

 
\end{abstract}

\section{Introduction}

Many structures we observe in the Universe have been shaped by fluid 
turbulence. In astronomy, we often observe high speed turbulence 
driven by supersonic ordered motions such as jets, supernova shocks 
and stellar winds (e.g. Franco \& Carraminana 1999). Hypersonic speeds, 
with Mach numbers above 10, are commonly encountered. Clearly, to 
understand the structure, we require a theory for supersonic turbulence. 
Here, we concentrate on decaying turbulence, such as could be expected 
in the wakes of bow shocks, in the lobes of radio galaxies or following 
explosive events. Two motivating questions are: how  fast does 
supersonic turbulence decay when not continuously replenished and how 
can we distinguish decaying turbulence from other dynamical forms? The 
first question has been answered through recent  numerical simulations 
described below. The answer to the second question is sought here. We 
look for a deep understanding of the dynamics and physics which control 
decaying supersonic turbulence. From this, and a following study of 
driven turbulence, we can derive the analytical characteristics and 
the observational signatures pertaining to supersonic turbulence. 
We caution that we specify to uniform three dimensional turbulence with 
an isothermal equation of state, an initially uniform magnetic field 
and periodic boundary conditions.

Numerical studies of decaying supersonic turbulence in three dimensions 
have revealed a power-law decay of the energy in time following  a short 
low-loss period (Mac Low et al. 1998; Stone et al. 1998).
Simulations of decaying subsonic and incompressible turbulence show
similar temporal behaviour (e.g. Galtier et al. 1997), as 
discussed by Mac Low et al. (1999). In the numerical experiments, random 
Gaussian velocity fields were generated with small wavenumber 
disturbances. Magnetic fields were included of various strengths. 
Mac Low (1999) concluded that the decay is so rapid under all conditions 
that the motions  we observe in molecular clouds must be continuously 
driven. In this work, we analyse the Mach 5 simulations from Mac Low et 
al. (1998) as well as a new Mach 50 simulation. The hypersonic run 
should best illustrate the mechanisms behind the development and 
evolution of the shock field, possibly revealing asymptotic solutions.

The major goal is to derive the spectrum of shocks (the Shock 
Probability Distribution Function) generated by turbulence. Shocks 
are often responsible for  detailed bright features, such as 
filamentary and sheet structures, within which particles are highly 
excited. An example of a region which appears to contain  a  chaotic 
mixture of shocks, termed a 'Supersonic Turbulent Reactor' is the 
DR\,21 molecular hydrogen outflow, driven by a collimated wind 
from a high mass young star (Smith et al. 1998). The shock spectrum 
is related to the molecular excitation, with weak shocks being 
responsible for rotational excitation and strong shocks for
vibrational excitation.

Previous studies of compressible turbulence have concentrated on the 
density and velocity structure of the cold gas rather than the shocks. 
Three dimensional subsonic and transonic simulations (e.g. Porter 
et al. 1994; Falgarone et al. 1994), two dimensional supersonic motions 
(V\'azquez-Semadeni 1994) as well as three dimensional supersonic 
turbulence have been discussed (V\'azquez-Semadeni et al. 1996, Padoan 
et al. 1998). One attempts to describe and  fit the density and 
velocity structures observed in molecular clouds. This is often 
appropriate for the interpretation  of clouds since, although the Mach 
number is still high, the shock speeds are too low to produce bright 
features. The simulations analysed here are also being interpreted by 
Mac Low \& Ossenkopf (2000) in terms of density structure.

Despite a diversity of theory, and an increase in analytical knowledge,
a succinct understanding of turbulence has not been attained (see
Lesieur 1997). Therefore, we need not apologise for not fully 
interpreting the results for the supersonic case. We do not look for a 
Kolmogorov-inspired theory for two reasons. First, fully developed 
turbulence becomes increasingly non-Gaussian towards small scales. 
These intermittency effects dominate the statistics of velocity jumps 
in supersonic turbulence. Second: the strong compressibility implies 
that a wavenumber analysis is irrelevant since the energy spectrum of 
a shock or of a system of shocks is simply $k^{-2}$ (e.g. Gotoh \& 
Kraichnan 1993). Note that the Kolmogorov-like spectra found by Porter 
et al. (1994) appeared at late times when the flow is clearly subsonic 
(and also note that even the initial RMS Mach number was only unity, 
which rather stretches the definition of supersonic turbulence).

We attempt here to construct a physical model to describe the non-Gaussian 
Probability Density Functions (PDFs) for the shock waves. We  adapt 
the mapping closure analysis, as applied to Navier-Stokes 
(incompressible) and Burgers (one dimensional and pressure free) turbulence 
(Kraichnan 1990), to compressible turbulence. Phenomenological 
approaches, such as multifractal models or the log-normal model, 
are avoided since they have limited connection to the underlying physical 
mechanisms. In contrast, mapping closure follows the stretching and 
squeezing of the fluid, and the competition between  ram pressure, 
viscosity and advection determines the spectral form.

We study here compressible turbulence without gravity, self-gravity
or thermal conduction. No physical viscosity is modelled, but  numerical 
viscosity remains present, and  an artificial viscosity determines the 
dissipation in regions of strong  convergence. By strong convergence, we mean
high negative divergence of the velocity field, which thus 
correspond to the shock zones as shown in Fig.\,\ref{shockfield}.
Periodic boundary conditions were chosen for the finite difference 
ZEUS code simulations, fully described by Mac Low et al. (1998).  

The ZEUS code itself is a time-explicit second-order accurate 
finite difference code (Stone \& Norman 1992a,b). It is ideal for 
problems involving supersonic flow and is versatile, robust and 
well-tested. Although higher order codes are potentially more 
accurate, the high speed of the algorithms means that large 
problems can be solved at high resolution. This enables us to 
test for convergence of the energy dissipation rate (Mac Low 
et al. 1998), shock distributions (Sect. 2.4) and numerical viscosity
(Sect. 2.6). Furthermore, the  basic hydrocode results have been confirmed on 
solving the same problems with the contrasting method of smoothed
particle hydrodynamics (Mac Low et al. 1998). The constrained transport 
algorithm (Evans \& Hawley 1988) updated through use of the method of 
characteristics (Hawley \& Stone 1995) is used to maintain a divergence-free 
magnetic field to machine accuracy and to properly upwind the advection.
 
We begin by discussing the method used to count shocks from grid-based 
simulations (Sect. 2.1-2.2). We then present the shock jump PDFs and provide 
analytical fits for the hypersonic M = 50 case (Sect. 2.3-2.5). The 
one-dimensional counting procedures are verified through a comparison 
with full three-dimensional integrations of the dissipated energy (Sect. 2.6). 
Supersonic hydrodynamic M = 5 (Sect. 3) and magnetohydrodynamic (Alf\'ven 
numbers A = 1 and A = 5) simulations (Sect. 4) are then then likewise 
explored. Note the definition of the  Alfv\'en Mach number 
$A = v_{\rm rms}/v_{\rm A}$, where $v_{\rm A}^2 = B^2 / 4 \pi \rho$ 
where $v_{\rm rms}$
is the initial root mean square (RMS) velocity and $v_{\rm A}$ is the
Alfv\'en speed. The evolution of the velocity PDFs are then presented and 
modelled (Sect. 5). Finally, we interpret the results in terms of the 
dynamical models (Sect. 6).

\section{Hydrodynamic hypersonic turbulence}
 
\subsection{Model description}

\begin{figure*}[hbt]
  \begin{center}
    \leavevmode
    \psfig{file=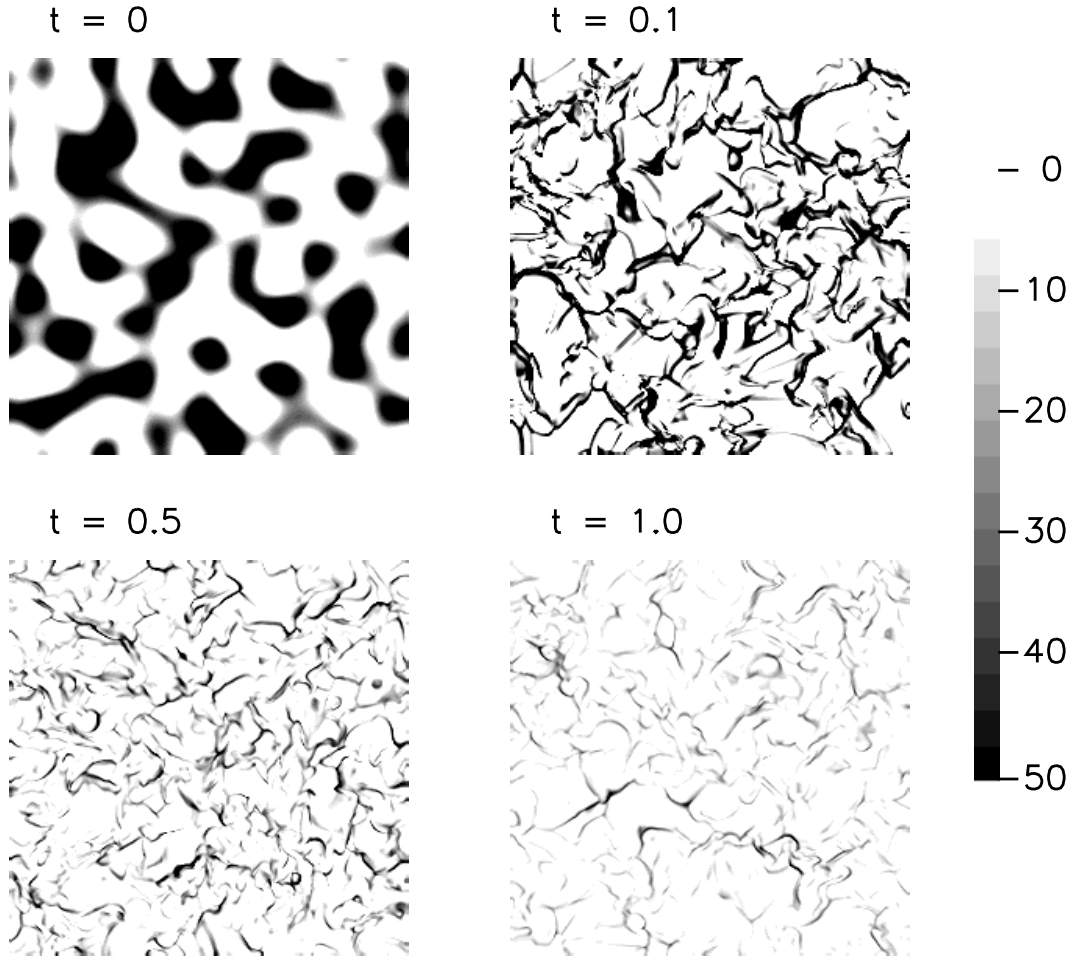,width=0.95\textwidth,angle=0}
\caption{Four stages in the early development of the shock field. The 
distribution of converging regions (i.e. the negative velocity divergence of 
the velocity field) is displayed here for a cross-sectional plane of 
the hydrodynamic M = 50 simulation. The displayed values are the 
numerical equivalents of the divergence (i.e. the sum of the velocity
differences) at each grid location. In order of increasing time, the 
minimum divergence (maximum convergence) and the average divergence 
within converging regions only in the displayed cross-sections are: 
(-178.4, -46.6), (-441.9, -48.0), (-146.7, -13.74) and (-99.9, -7.3).}   
\label{shockfield}
  \end{center}
\end{figure*}

The example we explore in detail is the decay of hypersonic hydrodynamic 
turbulence (Fig.\,\ref{shockfield}).  The three dimensional numerical 
simulation on a D$^3$ = 256$^3$ grid with periodic boundary conditions 
began with a root mean square Mach number of M = 50. The initial 
density is uniform and the initial velocity perturbations were drawn 
from a Gaussian random field, as described by Mac Low et al. (1998). 
The power spectrum of the perturbations is flat and limited to the 
wavenumber range $1 < k <  k_{max}$ with $k_{max} = 8$.

The simulations employ a box of length 2L (hence L is equivalent to
128 zones  in this example) and a unit of speed, $u$. The gas is 
isothermal with a sound speed 
of $c_s = 0.1u$. Hence the  sound crossing time is 20L/u. We take the time 
unit as $\tau$ = L/u. Thus a wave of Mach 50 would laterally cross the box 
in a time 0.4$\tau$ if unimpeded.

Cross-sections of the shock distribution are shown at four times in
Fig.\,\ref{shockfield}. We actually display a greyscale image,
biased to the high-convergence regions.  An evolution is hardly 
discernable in snapshots showing  all converging regions equally
shaded (as displayed in Fig. 1 of Mac Low et al. 1999). 

\subsection{Shock number distribution}

Ideally, we would like to calculate the total surface area for each 
shock strength as a function of time. We introduce the shock number
distribution function $dN/dv$, which is the number of shock elements per unit 
shock speed as a function of time and shock speed. A shock
element is the surface area of a shock put into units of the grid cell area.

To simplify our numerical analysis, we  calculate
instead the one dimensional shock jump function.
This is the number distribution  of the total jump in speed across
each converging region along a specific direction. This is written as
$dN/dv_j$ where $v_j$ is the sum of the (negative) velocity gradients
(i.e. $\Sigma\left[-{\delta}v_x\right]$ across a region being
compressed in the x-direction).
We employ the jump Mach number in the x-direction M$_j = v_j/c_s$ rather
than $v_j$ since this is the parameter relevant to  the  
dynamics. Thus, each bounded region of convergence in the x-direction
counts as a single shock and the total jump in M$_j$ across this region
is taken as its strength. 

Numerically, over the whole  simulation grid 
(x,y,z), we calculate each shock jump through
\begin{equation}
M_j =  \sum_{x=x_i}^{x=x_f} ({\Delta}v_x/c_s)
\end{equation}
with the condition that ${\Delta}v_x < 0$ in the range $x_i \le x < x_f$.
This is then binned  as a single shock element.
The shock number distribution $dN/dM_j$ is obviously dimensionless.

The 1D approach  neglects both the shock angle and full shock 
strength. The distribution of shock jumps, however, is  found by 
adding up an enormous number of contributions over the whole grid. 
This method has the advantage of being extremely robust, involving no model 
assumptions. To be a direct representation of the true shock 
strength function, however, it requires a few assumptions to be 
justified: (1) a one-dimensional shock jump is
 related to the actual shock speed, 
(2) not too many shocks are excluded because their surfaces 
are aligned with the chosen direction, (3) the shock velocities are
distributed isotropically and (4) unsteepened compressional waves
can be distinguished from true shocks. 

First, we note that it is an extremely difficult task to calculate 
the actual shock speed for each shock. It is, however, unnecessary 
since the shock speed and  one-dimensional shock jump are closely related  
statistically. We also take the number of zones at which  
compressive jumps are initiated as the number of shocks (where 
shocks are colliding, the method will be inaccurate). 
Assumption (3) will not hold for the magnetohydrodynamic 
turbulence which has a defined direction. In these cases, the jump 
distributions must be calculated both parallel and transverse to the 
original magnetic field. Assumption (4) will not be made: we include 
all acoustic waves, but we have followed the
width of the jumps and so can verify whether shocks or waves
are being counted. This is important since broad compressional waves 
also dissipate energy and become increasingly significant, of course, 
as the high-speed shocks decay and the flow eventually becomes subsonic.

Many shock surfaces are distorted, occasionally bow-shaped. This does 
not negate our counting procedure provided the curvature is not too 
strong. Here the relevant lengths are the shock radius of curvature 
and the shock width. The latter is determined by our numerical method, 
involving von Neumann \& Richtmyer (1950) artificial viscosity, which 
here constrains strong shocks to just a few zones. As seen from 
Fig.\,\ref{shockfield}, we can confidently take one-dimensional cuts 
across the shock surfaces and equate the measured jump to the
actual  jump in speed of fluid elements to a good first approximation.

In Sect.\,2.6, we check our method by demanding consistency with
integrated quantities derived directly from the numerical simulations.

\subsection{Hypersonic turbulence}

The random Gaussian  field  rapidly transforms into a shock field in the 
Mach 50 case (Fig.\,\ref{shockfield}). The shock steepening is reflected 
in the initial increase in the minimum value of the divergence (see the 
caption to Fig.\,\ref{shockfield}). Note that the {\em average} value of 
the divergence does not change i.e. the total number of converging zones 
only falls from half to about one third, despite the steepening. The 
explanation is that the shocks have time to form in the strongly 
converging regions but the compression in most of the flow progresses 
slowly. After the time t = 0.1, the number of fast shocks decays and  
the average convergence decreases. The total number of zones with 
convergence, however, remains constant  throughout. This fact, that the 
total shock surface area is roughly conserved,  is verified in the 
following analysis. 

\begin{figure*}
  \begin{center}
    \leavevmode
    \psfig{file=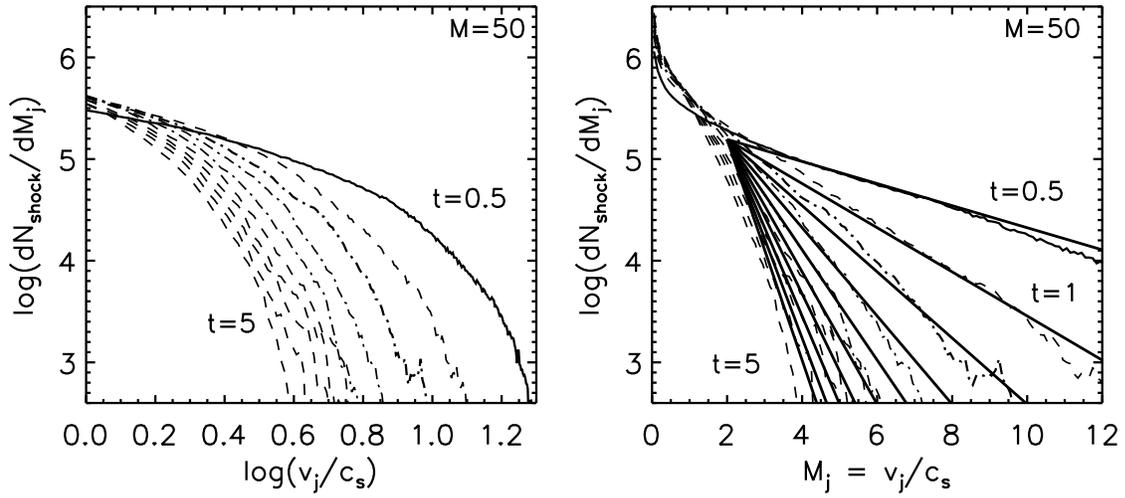,width=0.95\textwidth,angle=0}
\caption{The jump velocity distribution. Two representations of the
number of compressive layers in the x-direction as a function
of the total x-speed change across the layer at 10 equally-spaced times
beginning at time $t = 0.5$ (solid line) and continuing, with a
monotonic decay of the high-speed jumps, in time steps of 0.5. 
They demonstrate the decay and steepening and that (a) no power law can
represent the behaviour at any stage, and (b) a series of exponentials of
the form ${\exp}(-t(v_j-2)/2)$ fit the high velocities well. }   
\label{number}
  \end{center}
\end{figure*}

The one-dimensional distribution of the number of shock jumps as a
function of time is presented in Fig.\,\ref{number} for the case with 
RMS Mach number M = 50. This demonstrates that the shock jump function 
both decays and steepens. One can contrast this to the decay of
incompressible turbulence where the distribution function, as
measured by wavenumber, maintains the canonical Kolmogorov power-law in 
the inertial range during the decay (Lesieur 1997). Here we remark that 
a power-law fit is impossible (Fig.\,\ref{number}a) but stress that this 
result applies only to the case at hand: {\em decaying} turbulence.

\begin{figure*}
  \begin{center}
    \leavevmode
    \psfig{file=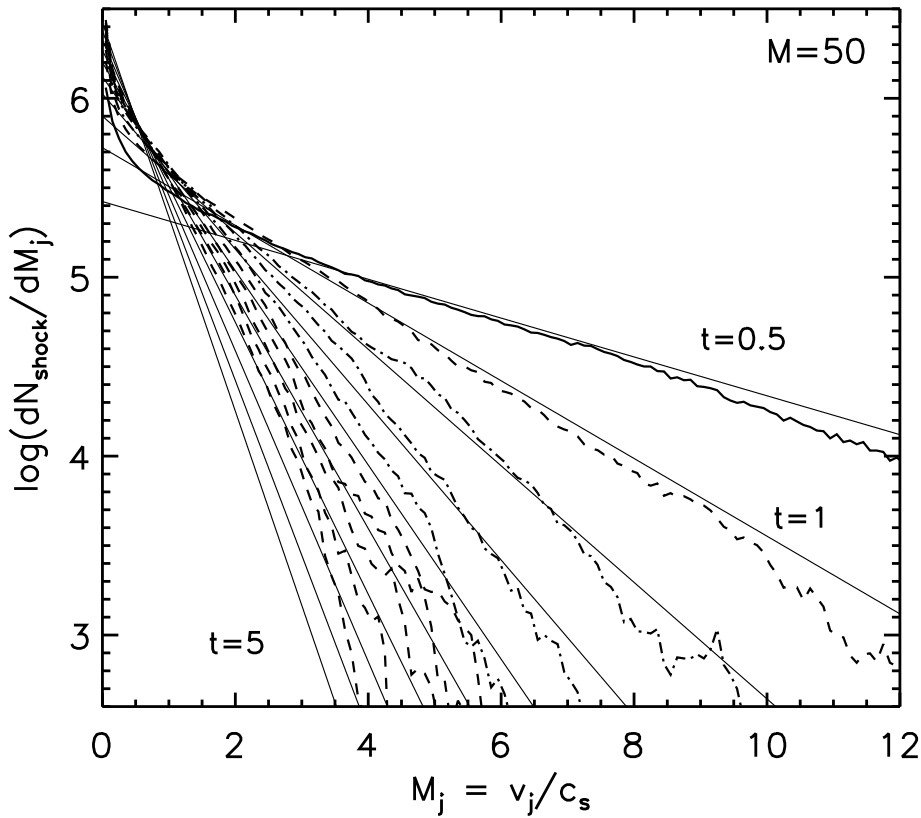,width=0.95\textwidth,angle=0}
\caption{The jump velocity distribution extracted from the M = 50
hydrodynamic simulation as a function of time, as in Fig.\,\ref{number}.
The fitted function is $dN/dM_j\,=\,10^{5.42}\,t\,{\exp}(-tM_j/2)$. }   
\label{fitted} 
  \end{center}
\end{figure*}

The jump distribution is very close to being exponential in both
velocity and time. This remarkably simple conclusion is based on the
good fits shown in Fig.\,\ref{number}b. Note that the
pure exponential only applies to the medium and strong shock regime.
To also account for the low Mach number regime, we fitted a further time
dependence as shown in Fig.\,\ref{fitted}, yielding
\begin{equation}
\frac{dN}{dM_j} = 10^{5.72}\,t\,\exp (-M_j\,t/2.0)
\end{equation}
in terms of the 1-D jump Mach number. Better fits can be obtained
with a somewhat more complex time dependence. We find excellent fits for
\begin{equation} 
\frac{dN}{dM_j} = 10^{5.79}\,t\,{\exp}\left[-\,{\beta}\,M_j\,t^{\alpha}\right]
\end{equation}
with $\alpha \sim 0.88 \pm 0.03$ and $\beta \sim 0.52 \pm 0.03$. The
values and errors are derived from parameter fitting of
 all the displayed curves to within a factor of $\sim 1.5$.
We exclude in this process the phase where collisional
equilibrium would not be expected: at early times and low Mach 
numbers $t\,M_j < 0.5$.  Also, we exclude the jump speeds
where the jump counts are low
(shock numbers ${dN}/{dM_j} < 3000$). 

\subsection{Convergence and dependence on initial conditions}

Resolution studies are essential to confirm numerical results. One hopes 
that the results demonstrate convergence. This is plausible for supersonic 
flow in which the decay does not depend on the details of the viscosity 
or the details of the shock transitions. This has been confirmed for the
analysis of the total energy (Mac Low et al. 1998).

We compare available simulations for the hypersonic study with  64$^3$ 
and 128$^3$. We also set the initial wavenumber range to $k_{max} = 2$ 
and can thus examine the dependence on the chosen initial state.

\begin{figure}[t]
   \begin{center}
         \leavevmode
\psfig{file=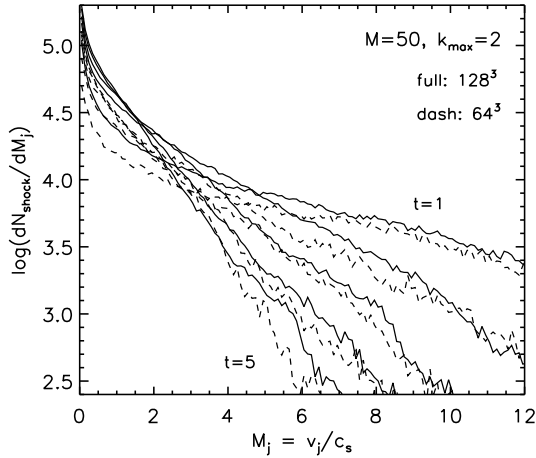,width=0.53\textwidth,angle=0}
\caption{The evolution of the jump velocity distribution at two 
resolutions for M = 50 hydrodynamic simulation with k$_{max}$ = 2. The
shock number for the 64$^3$ simulation has been multiplied by 4 to
adjust for the larger zone sizes. }   
         \label{number2} 
    \end{center}
\end{figure}

\begin{figure*}
  \begin{center}
    \leavevmode
    \psfig{file=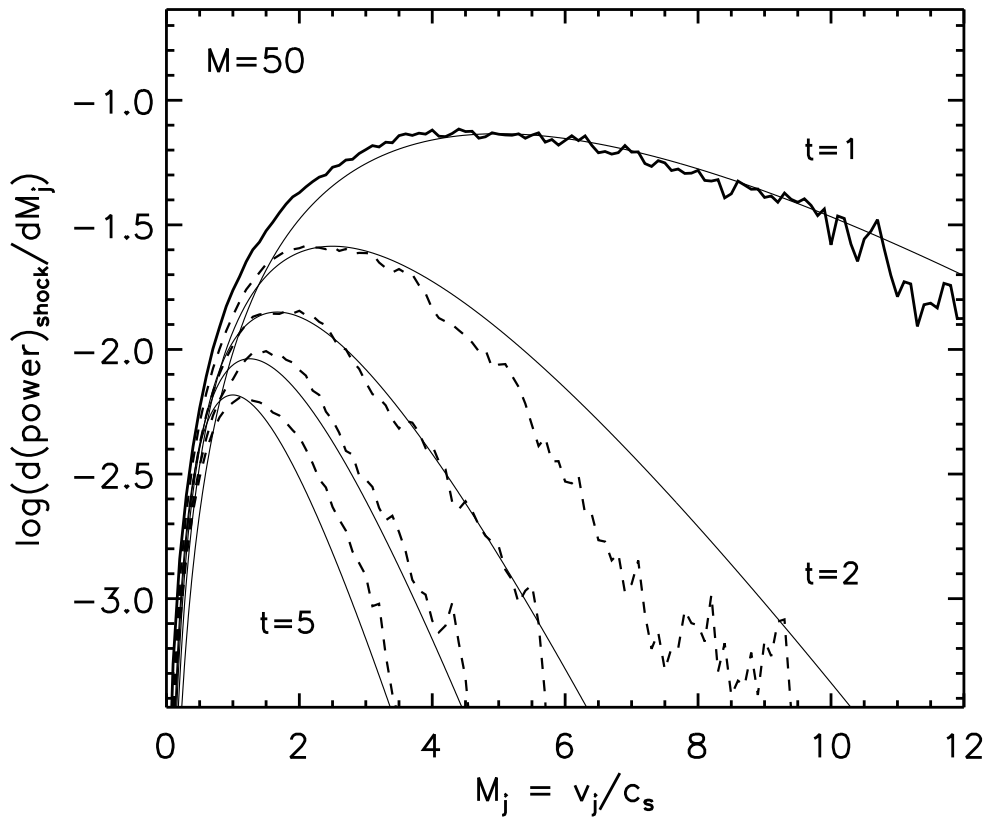,width=0.95\textwidth,angle=0}
\caption{The power dissipated as a function of the jump velocity. The 
power is displayed per unit M$_j$ where M$_j$ = v$_j$/c$_s$, and the 
data is extracted from the M = 50 hydrodynamic simulation for times 
t = 1,2,3,4 \& 5. The fitted function is 
$d\dot E/dM_j\,=\,0.016\,t\,M_j^{2.5}\,{\exp}(-tM_j/2)$.}   
\label{power50}
  \end{center}
\end{figure*}

The results at the two different resolutions are in quite good agreement,
especially in the high Mach number regime (Fig.\,\ref{number2}). The density 
of high Mach number
shocks is quite low and they are well resolved.  At low Mach numbers, the lower
resolution simulation fails, of course, to capture the vast quantities
of weak compressional waves contained in the higher resolution example.
It is to be expected that shock turbulence configurations get 
extremely complex on small scales, through the interactions which produce
triple-point and slip-layer structures. To capture this structure
requires adaptive grid codes. This does not mean, however, that
the simulations are inaccurate for our purposes since energy dissipation is not
controlled by the weak shocks until very late times, as verified in
Fig.\,\ref{power50}. 

\begin{figure}[hbt]
  \begin{center}
    \leavevmode
    \psfig{file=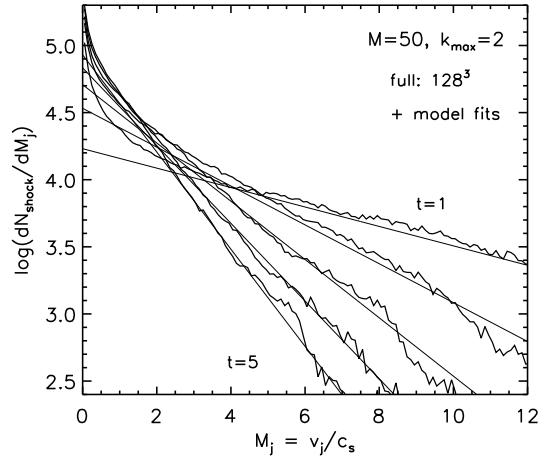,width=0.53\textwidth,angle=0}
\caption{The jump velocity distribution extracted from the 128$^3$ M = 50
hydrodynamic simulation with k$_{max}$ = 2 as a function of time.
The fitted function is $dN/dM_j\,=\,10^{3.93}\,t\,{\exp}(-tM_j/6)$. }   
\label{fitted2} 
  \end{center}
\end{figure}

A similar formula for the shock jump function is found. The evolution, 
however, is three times slower. We show in Fig.\,\ref{fitted2} the model fit
\begin{equation}
\frac{dN}{dM_j} = 10^{4.23}\,t\,\exp (-M_j\,t/6.0).
\end{equation}

This suggests that the rate of decay is proportional to the initial 
mean wavenumber of the wave distribution. The decay $k_{max} = 2$ 
simulation is a factor of three slower than in the $k_{max} = 8$ 
simulation. The mean wavenumber of the two simulations are $k_m = 1.5$ 
and $4.5$, with flat initial energy distributions. Hence, given that 
$M_j = v_x/c_s$ and t is in units of $L/10c_s$, the shock numbers
are approximately  $\propto t\,{\exp}(-k_ot\,v_j)$ where
$k_o \sim 1.1\,k_m$. The dependence on the initial mean wavenumber is
expected since the box size should not influence the decay rate if
we are indeed, as we wish, following the unbounded decay.

Note that two parameter fitting, as above, in this case yields
\begin{equation} 
          \frac{dN}{dM_j} = 
          10^{4.23}\,t\,{\exp}\left[-\,{\beta}\,M_j\,t^{\alpha}\right]
\end{equation}
with $\alpha \sim 1.02 \pm 0.03$ and $\beta \sim 0.168 \pm 0.008$.

\subsection{Shock power distribution}

How is the spectrum of shock jumps related to the decay of energy? Here
we show that the energy dissipation in the fast shocks is directly
correlated with their number which is decreasing
exponentially.  Furthermore, the weakest shocks, which merge into
an area of 'compressional waves', are ineffective in the overall
dissipation. The result is that the  moderately-supersonic part
of the turbulence rapidly becomes and remains responsible for the 
energy dissipation for an extended time. 

The shock power distribution function  is here defined as the energy
dissipated per unit time per unit jump speed $v_j$ as a function of
jump speed. Here again we employ the uni-directional jump Mach number M$_j$.
We actually calculate the energy dissipated by artificial viscosity
acting along a specific direction  within the shocks as defined by 
convergence along this direction. Hence we anticipate that in isotropic 
turbulence one third of the full loss will be obtained. The relative 
contributions of artificial and  numerical viscosity, which also confirms 
the method employed here, are discussed in Sect.~2.5.

The jump {\em number} distribution  includes a high proportion of very 
weak compressional waves that dissipate little energy. The 
one-dimensional shock power distribution, d$\dot E$/dM$_j$, shown in 
Fig.\,\ref{power50}, illustrates this. The functional fit is guided by 
the above shock number distributions, which we would expect to remain 
accurate for the high Mach number jumps. We display the fit to:
\begin{equation}
          \frac{d\dot E}{dM_j} =  0.016\,M_j^{2.5}\,t\, \exp(-k\,M_j\,t)
\label{eqnpow}
\end{equation}
with k = 0.5, which is again remarkably accurate given the lack of 
adjustable parameters.
Hence ${d\dot E}/{dM_j} \propto M_j^{2.5}{d\dot N}/{dM_j}$. Note that the
energy dissipated across an isothermal steady shock of Mach number M and 
pre-shock density
$\rho$ is $\dot E_s$ = (${\rho}c_s^3$/2)\,M$^3$(1\,-\,1/M$^2$) and that 
the jump in Mach number is M$_J$ = M(1\,-\,1/M$^2$). This yields
\begin{equation}
        \dot E_s = 0.5\,{\rho}c_s^3\,M_J^3\,
        \left[1\,+\frac{\surd(1+4/M_J^2)\,-\,1}{2}\right].
\end{equation}
Therefore, the numerical result suggests a (statistical) inverse
correlation between density and shock strength.

\begin{figure}[hbt]
  \begin{center}
    \leavevmode
    \psfig{file=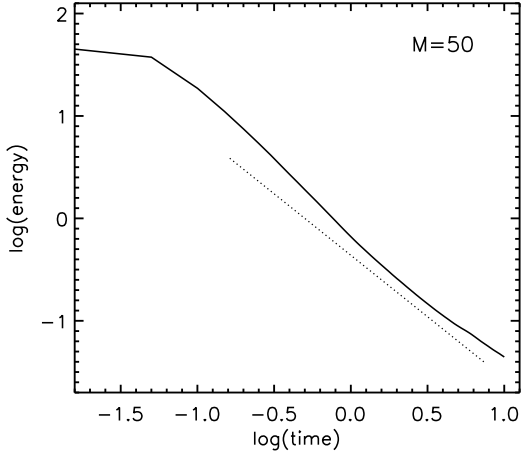,width=0.53\textwidth,angle=0}
\caption{The total energy as a function of time (full line) from data 
extracted from the M = 50 hydrodynamic simulation. For comparison, the  
dashed line is a power law with exponent -1.5.}   
\label{powertime}
  \end{center}
\end{figure}

Note that Eq.\,\ref{eqnpow} yields $\dot E \propto t^{-2.5}$ (on 
integrating over $M_j$ and substituting the variable $w = M_jt$) 
Thus, one obtains a power-law decay in total energy of the form E 
$\propto$ t$^{-1.5}$. This  behaviour of the total energy decay and 
the energy dissipation rate, is indeed found in the simulations, 
as shown in Fig.\,\ref{powertime}. Hence the results are fully 
consistent with the simple fits.
Note that the energy decay deviates from a pure power law: there is 
an early transition phase between times t = 0.1 - 0.5 during which 
the decay is more rapid (Fig.\,\ref{powertime}). This was not found 
in simulations of moderate Mach number supersonic turbulence (Mac 
Low et al. 1998) but appears here and also in simulations in which 
the supersonic turbulence is initially driven (Mac Low 1999). 

\subsection{Shock power and artificial viscosity}
In order to check the approximations described in Sect.~2.2, we analyse
how the energy dissipated due to artificial viscosity
varies over time and with different parameters of the model. For 
non-driven, or freely decaying runs, the following equation is 
taken as a basic:
\begin{equation}
     \frac{{dE}}{dt} = - (H_{\nu} + H_{numerical})\label{basic_h},
\end{equation}
where $H_{\nu}$ and $H_{numerical}$ are energy dissipation rates due
to artificial viscosity and numerical viscosity respectively. 

\begin{figure}
  \begin{center}
    \includegraphics[width=0.485\textwidth]{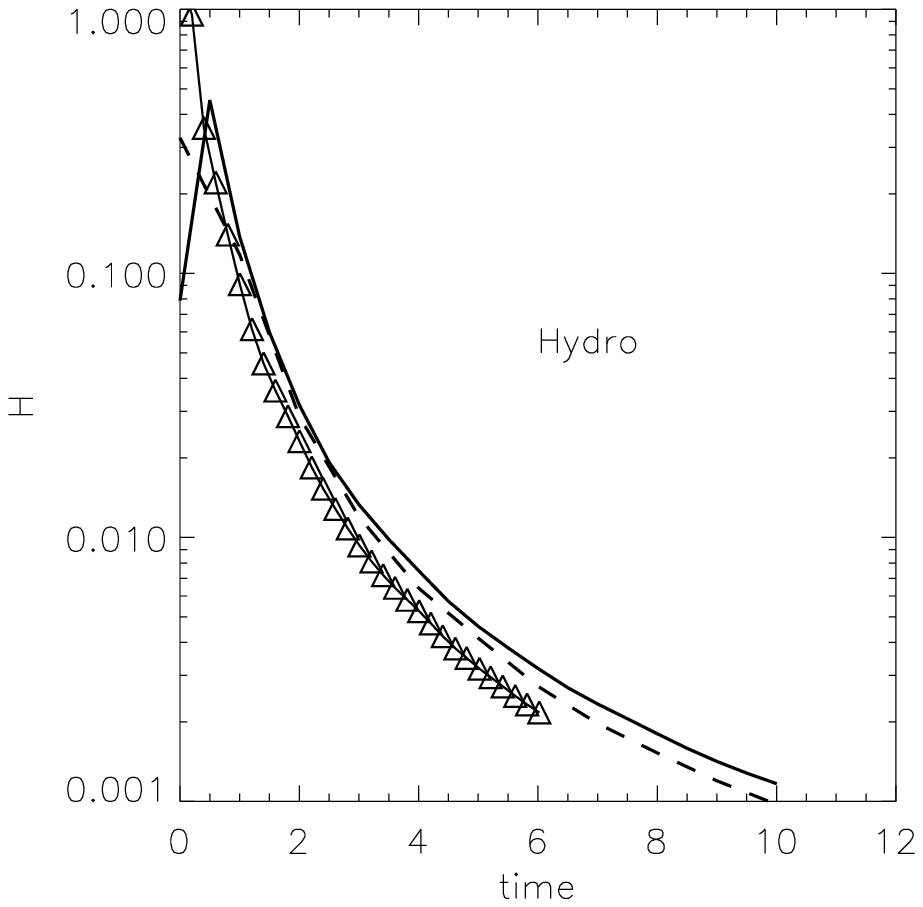}
    \includegraphics[width=0.485\textwidth]{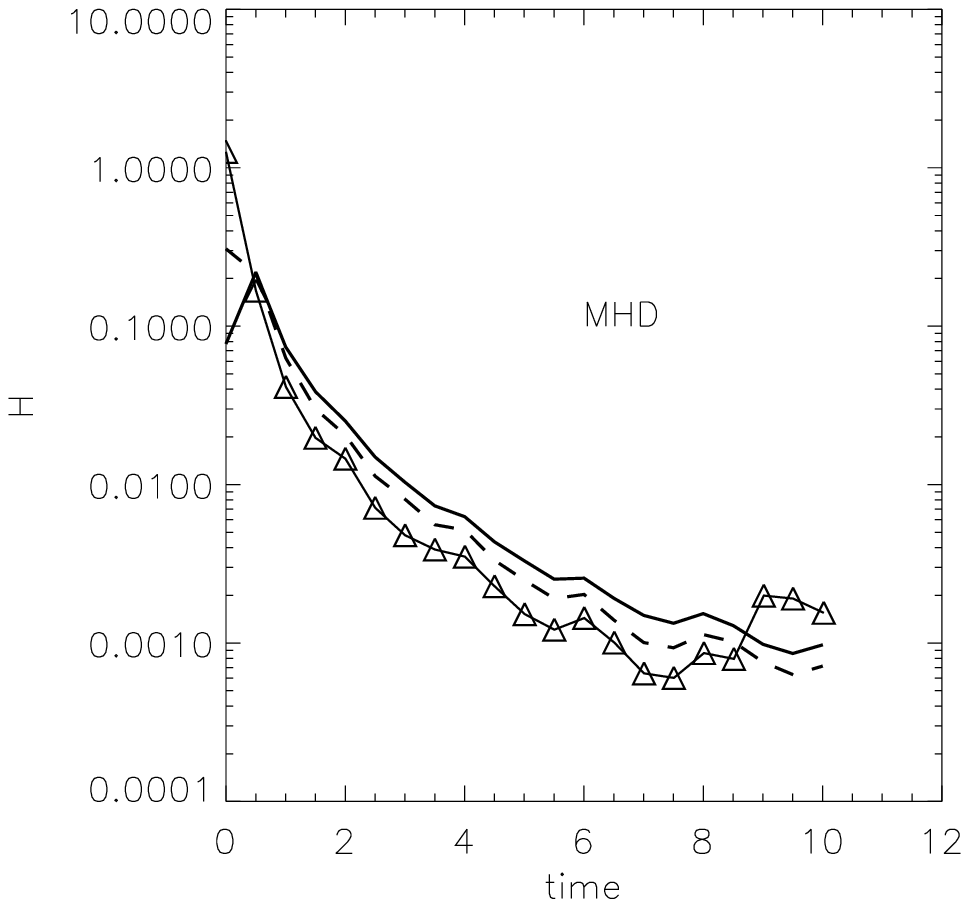}
     \end{center}
\caption{Resolution study for 3D models -- Energy dissipation rate due
to artificial viscosity versus time. Top graph: model B (64$^3$,
triangle), model C (128$^3$, dashed line), and model D (256$^3$, solid
line). Bottom graph: model N (64$^3$, triangle), model P (128$^3$,
dashed line), and model Q (256$^3$, solid line). For hydrodynamic
models we observe that the energy dissipation rate $H_{\nu}$ has
converged to better than 35\% moving from 64$^3$ to the 128$^3$
models, and to better than 25\% moving from the 128$^3$ to the 256$^3$
models. These values are 37\% and 23\% for the MHD models. Thus,
energy dissipation rate due to artificial viscosity converges as we go
to finer grids.}
\label{fig:g1}  
\end{figure}

We follow Stone \& Norman (1992) Eqs. (32)-(34), (134) in
computing $H_{\nu}$. The first step in computing derivatives is to realize that
artificial viscosity operates on compression only, so points with positive
derivatives in each direction are set to zero :
\begin{equation}
  {\partial{v1}}_{i,j,k} = \left\{ \begin{array}
   {l@{\quad \quad}l}
    0 & {\mbox{if   } (v1_{i+1,j,k}-v1_{i,j,k})> 0} \\
    {v1_{i+1,j,k}-v1_{i,j,k}} & {\mbox{otherwise}}.
  \end{array} \right.  
\end{equation}
Then scalar artificial pressures $q_i$ in all 3 directions are 
computed, with the uniform mesh ($\Delta x$) :
\begin{equation}
       q_i = {l^2}{\rho}{\left( \frac{\partial{v_i}}{\partial{x_i}} 
         \right)}^2 = {\left[ \frac{l}{\Delta x}\right] }^2 
         {\rho} {({\partial{v_i}})}^2
\end{equation}
where ${(l/{\Delta{x}})}^2$ is a dimensionless constant which
measures the number of zones over which the artificial viscosity will
spread a shock and was chosen to be 2 in these simulations. Then we 
calculate the artificial viscosity tensor $h_{\nu}$:
\begin{equation}
  {\stackrel{\longleftrightarrow}{h_{\nu}}} = -{\stackrel{\longleftrightarrow}{\nabla{v}}} : {\stackrel{\leftrightarrow}{Q}} = {(-1)}{\left[{\frac {\partial{v1}}{\partial {x1}}}{q1} + {\frac {\partial{v2}}{\partial {x2}}}{q2} + {\frac {\partial{v3}}{\partial {x3}}}{q3} \right]}  
\end{equation}
and compute the artificial viscosity dissipation rate for the entire 
cube at each particular time dump as 
\begin{equation}
      H_{\nu} = \int{\stackrel{\longleftrightarrow}{h_{\nu}}} dx^3   \sim
      \sum_{ijk} (h_{\nu,ijk}) \delta x^3.  
\end{equation}
To understand the convergence properties of the energy dissipation rate, 
we performed a resolution study for resolutions of $64^3$, $128^3$,
and $256^3$ zones for both hydrodynamic and MHD models.  Fig.~\ref{fig:g1} 
shows that the energy dissipation rate $H_{\nu}$ has converged to better 
than 25\% moving from the 128$^3$ to the 256$^3$ models.

Now let us examine the fraction of the total kinetic energy lost to
dissipation in shocks due to artificial viscosity $R = H_{\nu}/ (dE/dt)$.
We avoid taking this ratio locally in time as variations of $H_{\nu}$
in time around the average lead to spurious results. A more robust way 
to compute this ratio is to integrate over the entire curve, thus 
averaging over the variations in $H_{\nu}$ by computing
\begin{equation} \label{ratio}
R = \frac {\int H dt}{E_k(t_f) - E_k(t_0)}. 
\end{equation}
Table 1 shows these ratios for a set of models with $M=5$ and $k_{max}=8$. 
\begin{table*}
\begin{center}
\begin{tabular}{ccccccccc} \hline
\multicolumn{9}{c}{\bfseries Comparison of Ratios} \\ \hline
run & B &C & D & N & P & Q & J & L \\
\hline 
  grid &64$^3$ &128$^3$ &256$^3$ &64$^3$ &128$^3$ &256$^3$ &64$^3$ &256$^3$\\
  physics &Hydro &Hydro &Hydro &MHD &MHD &MHD &MHD &MHD\\
  A & $\infty$ & $\infty$ & $\infty$ &1 &1 &1 &5 &5\\
  $R$ &0.62 &0.68 &0.68 &0.38 &0.36 &0.35 &0.59 &0.49\\
  \hline \end{tabular}
  \caption{The fraction of the energy dissipated through artificial
  viscosity for models of supersonic turbulence.  Ratio $R$ is defined
  by Eq.~(\ref{ratio}). Model labels correspond to those of Table 1
  of Mac Low et al. (1998). A is the RMS Alfv\'en Mach number.}
\end{center}
\end{table*}
The table indicates that the energy dissipation ratio R has converged to 
a few percent in the hydrodynamic case at the resolution of 128$^3$, and 
even in the MHD case is converged to better than 3\% at $256^3$.

We find that the fraction of energy lost in shocks to artificial
viscosity doubles when we go from MHD to hydrodynamic models. Mac Low
et al. (1998) speculated that runs with magnetic fields dissipating 
much of their energy via short-wavelength MHD waves.  The factor of 
two difference in the dissipation rate due to artificial viscosity from
hydrodynamic to MHD runs gives further evidence for this dissipation 
mechanism. This behavior appears well-converged, as discussed above.

\section{Supersonic turbulence: M = 5}

Moderate-speed hydrodynamic turbulence has been discussed by 
Mac Low et al. (1998). They also  found that the kinetic energy decreases
with time as a power law, but with a shallower exponent. For the M=5
study (Model C) they recovered a E $\propto t^{-1.0}$ law. 

\begin{figure*}
  \begin{center}
    \leavevmode
    \psfig{file=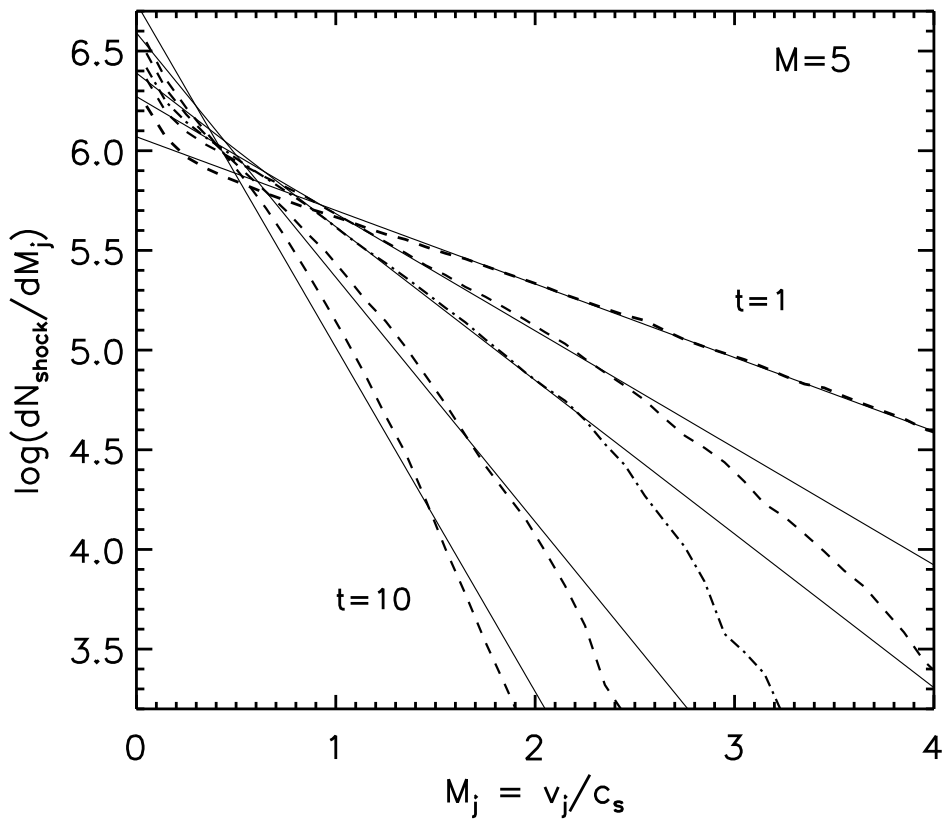,width=0.95\textwidth,angle=0}
\caption{The jump velocity distribution extracted from the M = 5
hydrodynamic simulation (Mac Low et al. 1998) as a function of time.
The fitted function is exponential in velocity but more complicated
in time (see text) ($\alpha = 0.67$ and $\beta = 0.85$ displayed).
The time sequence shown is t = 1,2,3,5 and 10.}   
\label{number5} 
  \end{center}
\end{figure*}

\begin{figure*}
  \begin{center}
    \leavevmode
    \psfig{file=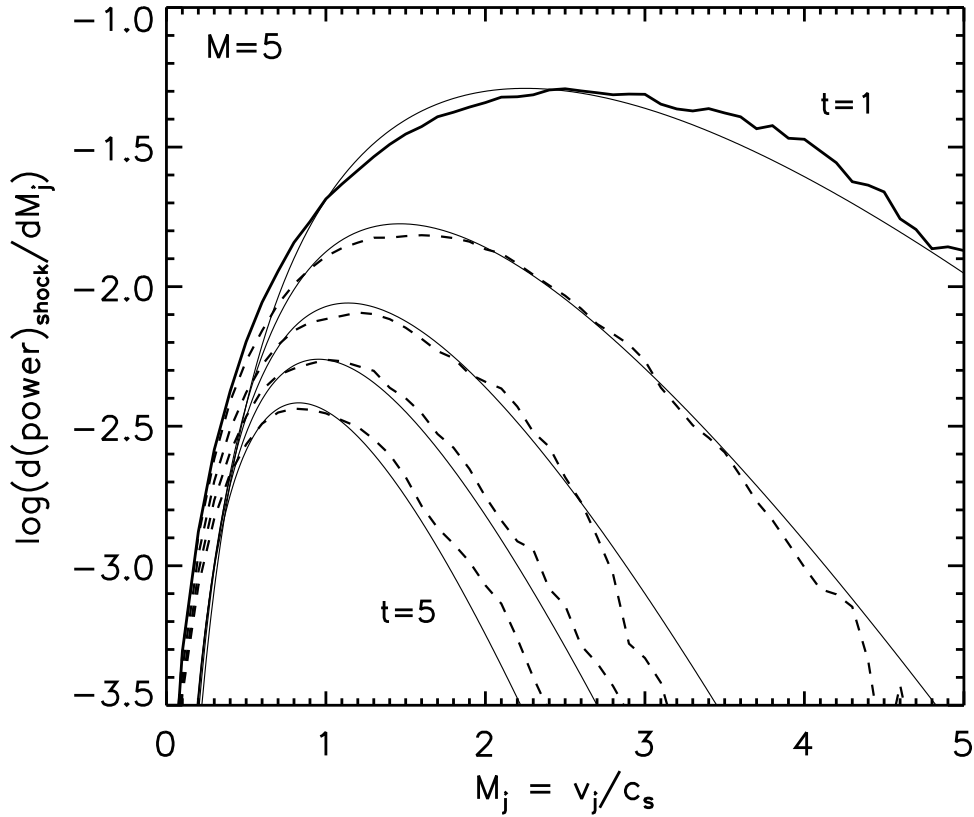,width=0.95\textwidth,angle=0}
    \caption{The power dissipated as a function of the jump
             velocity. The power is displayed per unit M$_j$
             where M$_j$ = v$_j$/c$_s$, and the data is extracted 
             from the M = 5 hydrodynamic simulation for times 
             t = 1,2,3,4 \& 5. The fitted function, given in the text,
             takes $\alpha = 0.62$ and $\beta = 1.6$.}   
    \label{power5}
  \end{center}
\end{figure*}

The shock jump distribution for M = 5 is shown in Fig.~\ref{number5}. 
Exponential velocity distributions are again found even though the
range in Mach numbers over which we could expect a specific
law is rather narrow (M$_j$ $\sim 1 - 3$). Pure exponential
time fits, however, are not accurate. We display a suitable 
fit, of the form
\begin{equation} 
\frac{dN}{dM_j} = 10^{6.07}\,t^{\alpha}\,{\exp}(-{\beta}\,M_jt^{\alpha})
\end{equation}
with $\alpha \sim 0.67 \pm 0.05$ and $\beta \sim 0.85 \pm 0.1$.
The corresponding  shock power distribution is shown in Fig.\,\ref{power5} 
along with a best-fit family of curves calculated from
\begin{equation} 
\frac{d\dot E}{dM_j} = 0.10\,M_j^{3.6}\,t^{\alpha}\,
                       {\exp}(-{\beta}\,M_jt^{\alpha})
\label{eqnpower}
\end{equation}
with $\alpha = 0.62 \pm 0.04$ and $\beta = 1.6 \pm 0.1$.Hence, an exponential velocity distribution is maintained. The decay,
however, is slightly slower. Integrating over $M_j$, yields
$E \propto t^{-1.00}$.

Integrating over $M_j$, with the limits of integration from 0 to $\infty$ (to
account for all the jumps), and with the substitution $x = M
t^{\alpha}$, we observe that the integral becomes
time-independent. Thus, integration of Eq.\,\ref{eqnpower} over $M_j$ yields
${dE/dt} \propto t^{-2.23}$. This yields $dE/dt \propto
t^{-1.23}$, which is quite close to the $E \propto t^{-1}$ 
law found by Mac Low et al.\ (1998). 

\begin{figure}
  \begin{center}
    \includegraphics[width=0.485\textwidth]{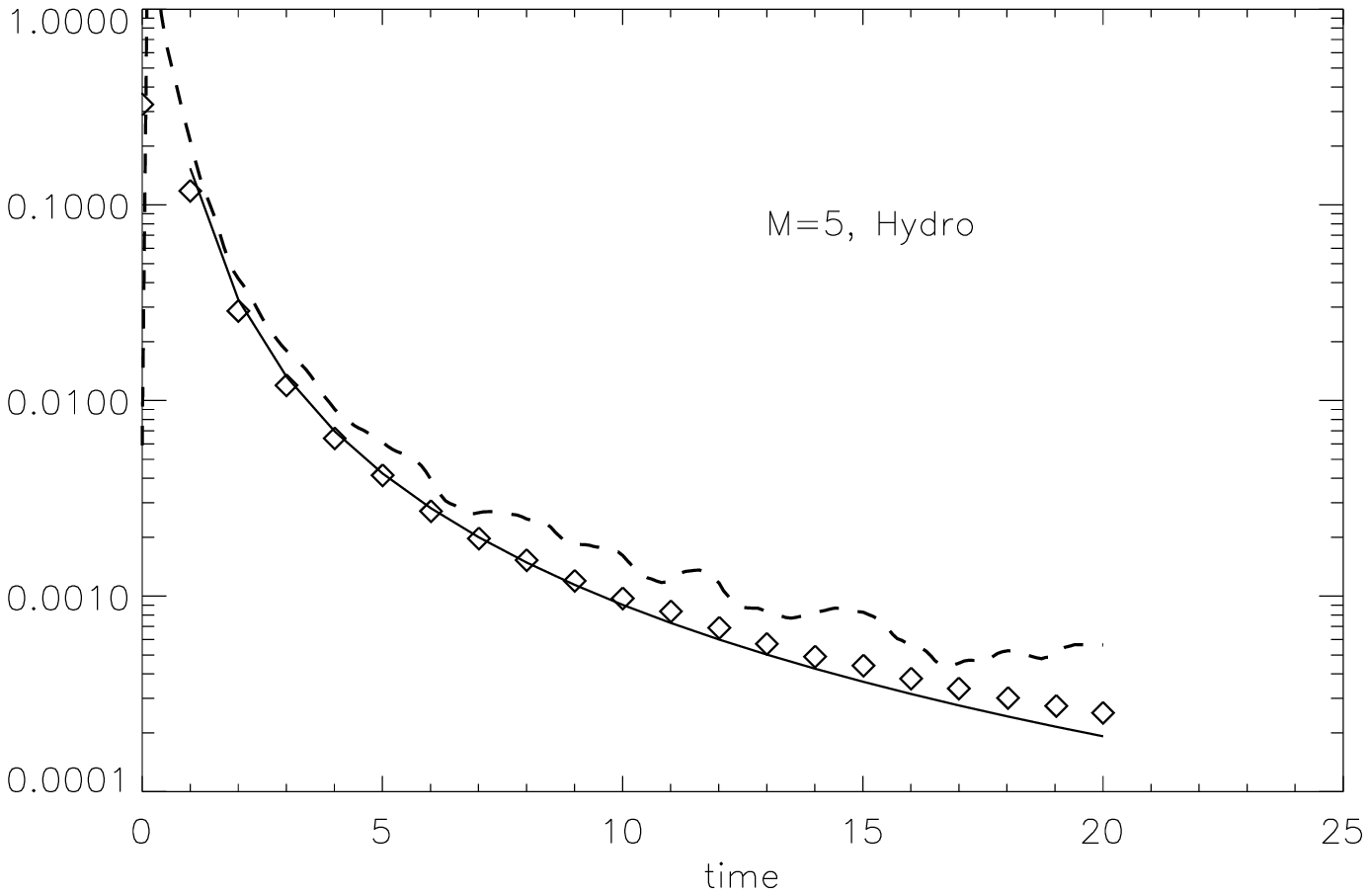}
    \includegraphics[width=0.485\textwidth]{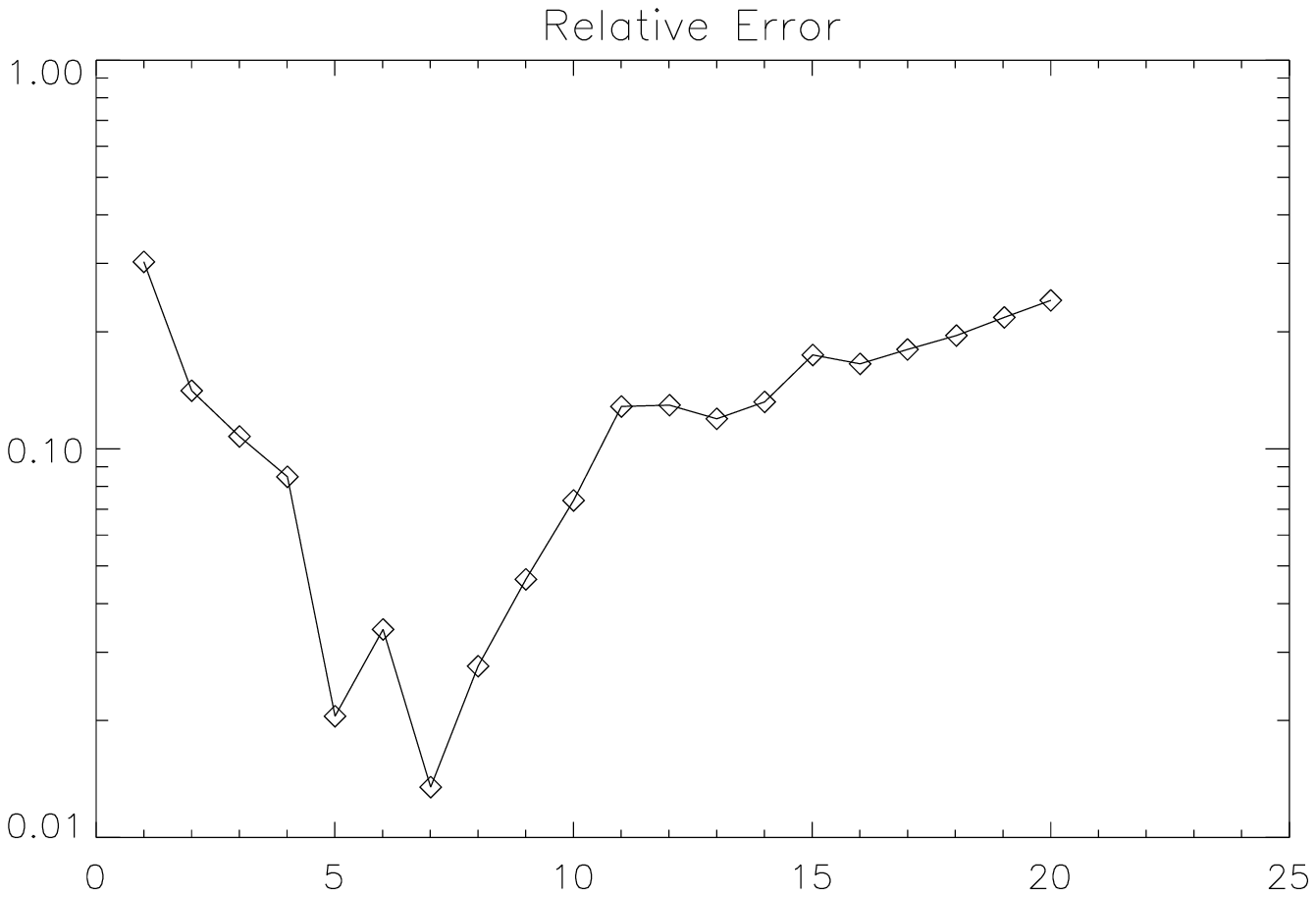} 
   \end{center}
\caption{ Top: dE/dt versus time - the graph presents $H_{\nu}$ for 
model C (grid 128, M=5, hydrodynamic) for 21 time dumps (rhombs), 
and integral of Eq. (8) over $dM_j$ (solid line). Both are energy 
dissipation rates due to artificial viscosity. We see a close fit of 
the methods. For comparison, dashed line represents total kinetic 
energy changes $dE/dt$ directly from the numerical simulation output. 
Bottom: the absolute value of the relative error between $H_{\nu}$ 
and integrated Eq. (\ref{eqnpower}).} 
\label{fig:f1}  
\end{figure}

Now we want to compare how the two quantities vary with time: $dE/dt$
from Eq.\,\ref{eqnpower} and the energy dissipation rate due to artificial 
viscosity $H_{\nu}$, derived using the algorithm
discussed in Sect. 2.6. Fig.~\ref{fig:f1} shows that
the two methods indeed agree. Thus, the shock power distribution
method (Sect. 2.6) confirms that the statistical approach works
with power calculations for each 1D converging region. 

\section{MHD turbulence: M = 5, A = 1 and 5}

An analysis of  simulations of  MHD turbulence allows us to
determine which  wave modes are involved.
The time dependence of the kinetic energy of freely decaying
MHD turbulence  has been discussed by 
Mac Low et al. (1998). Remarkably, the kinetic energy also decreases
with time as a power law, although with only a slightly shallower exponent
than the equivalent  hydrodynamic simulation. We consider here
the high-field example in which the initial RMS Mach number M = 5
and  the initial RMS Alfv\'en  number A
is unity  and the low-field equivalent with A = 5. 
Mac Low et al. recovered: E $\propto t^{-0.87}$.
(at the highest resolution of 256$^3$) for the high field case.

The initial field configuration is simply a uniform field aligned
with the z-axis. Thus the imposed velocity field controls the 
turbulent energy input, and some energy is subsequently transferred 
into magnetic waves. We impose no turbulent diffusion here: magnetic energy
may, however, still be lost through numerical diffusion or MHD wave 
processes.

\begin{figure}
  \begin{center}
    \leavevmode
    \psfig{file=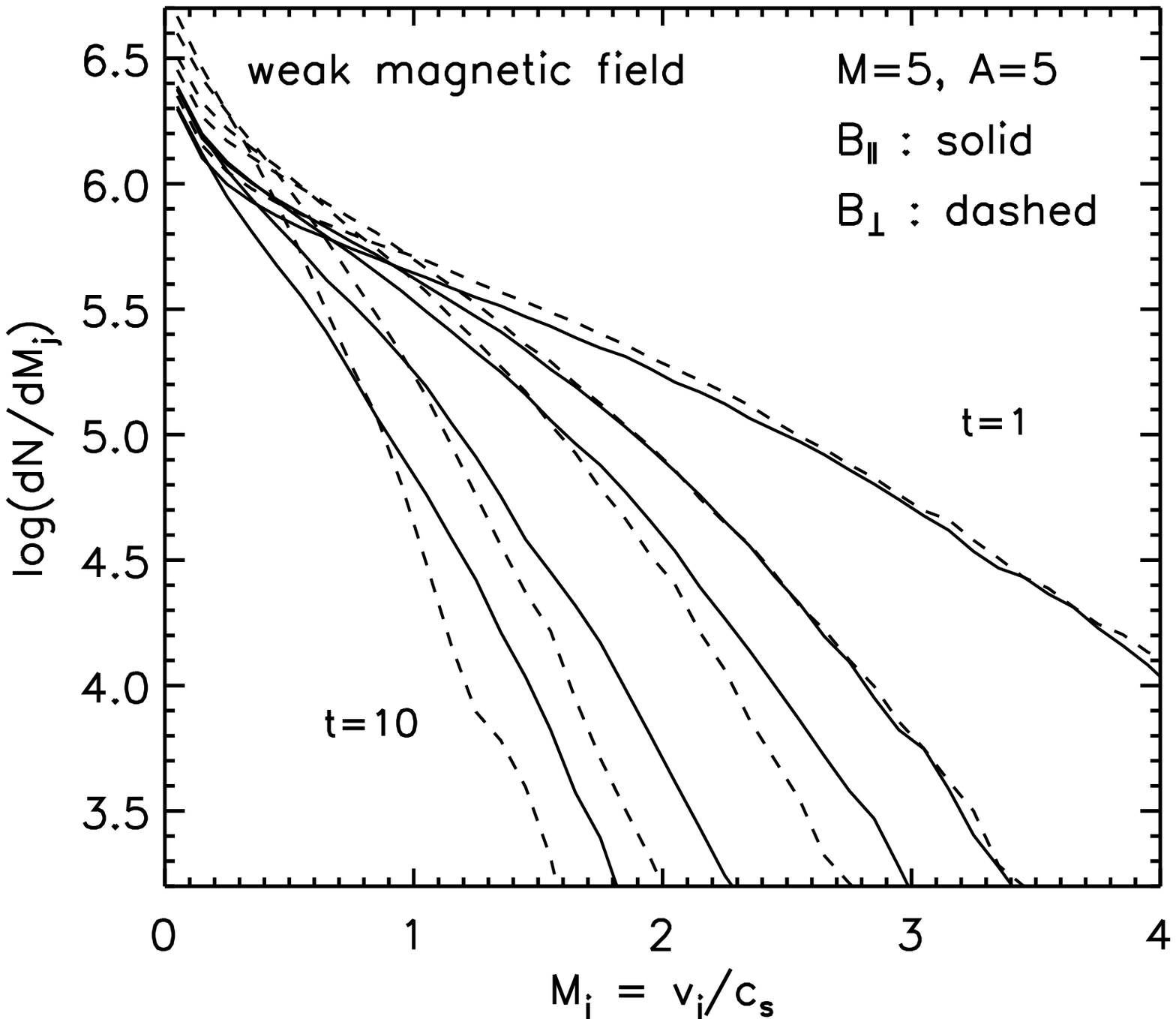,width=0.57\textwidth,angle=0}
    \psfig{file=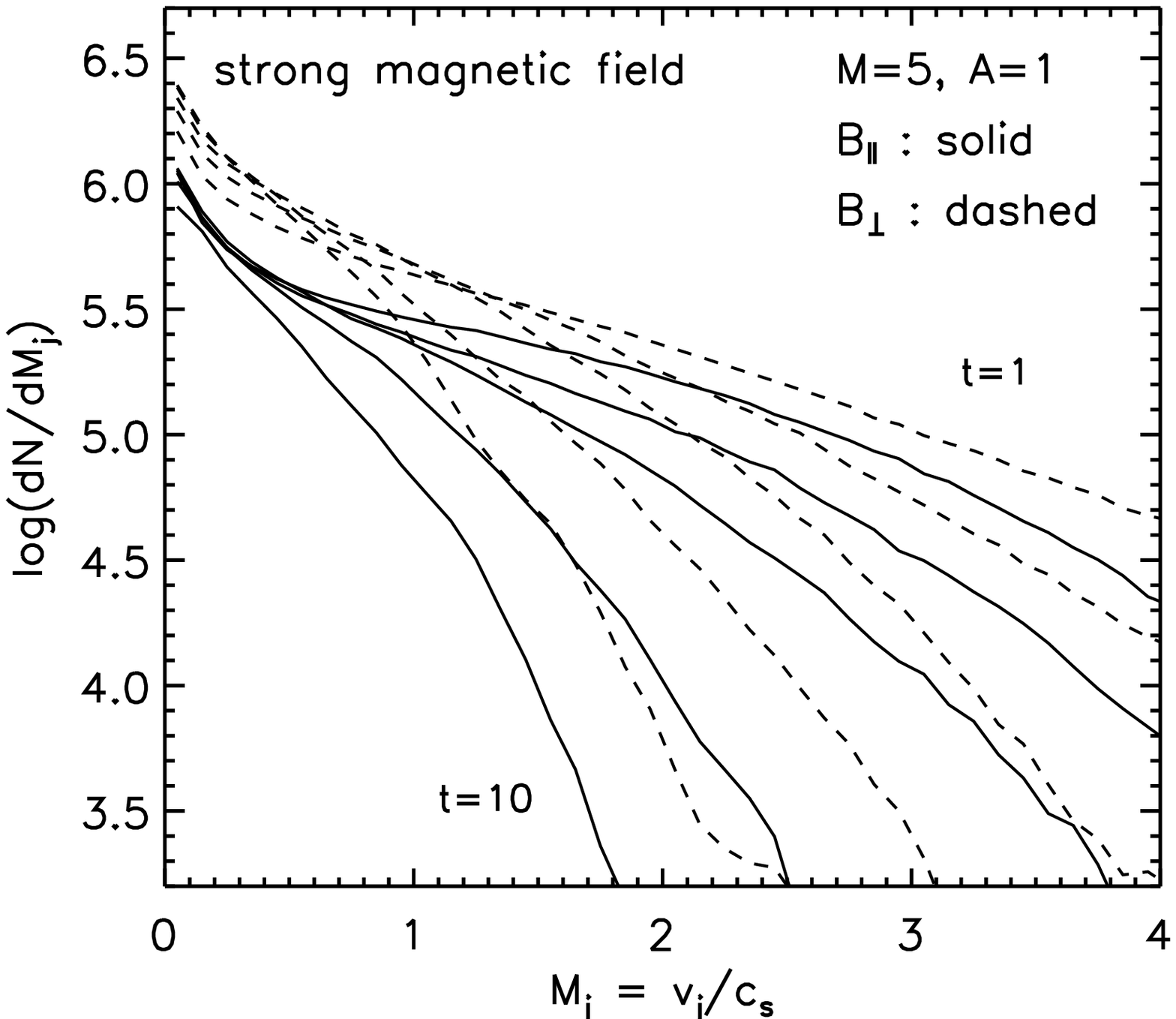,width=0.57\textwidth,angle=0}
\caption{The jump velocity distribution extracted from the M = 5
low magnetic field (A = 5) and high field (A = 1) MHD 256$^3$ 
simulations (Mac Low et al. 1998) as a function of time. The 5 
times shown are t = 1,2,3,6 and 10. The two curves shown for
each time are the velocity jumps transverse to the field (dashed) 
and parallel to the field (solid lines). The fitted exponentials 
are described in the text.}   
\label{numbermag} 
  \end{center}
\end{figure}

\begin{figure}
  \begin{center}
    \leavevmode
    \psfig{file=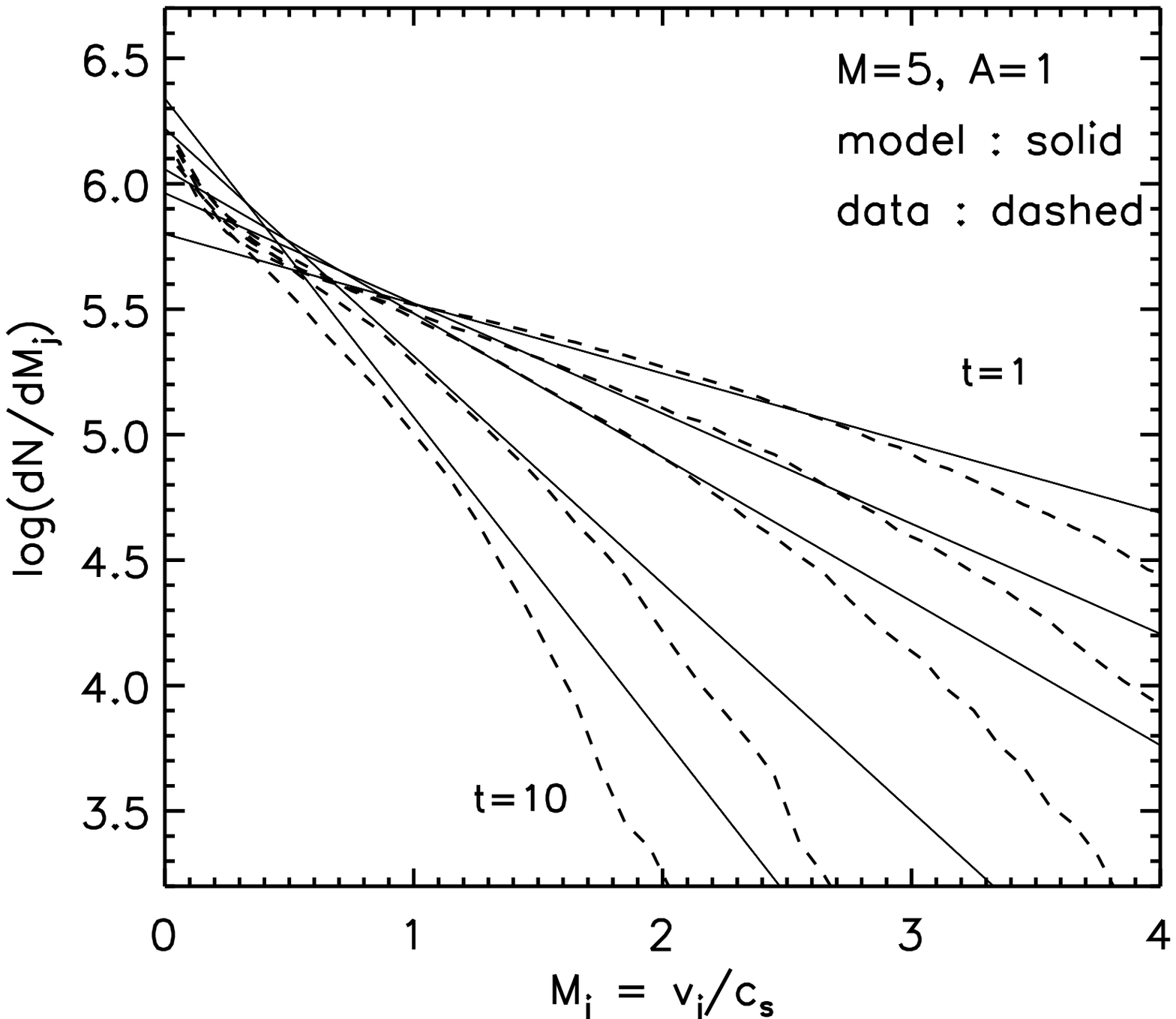,width=0.57\textwidth,angle=0}
\caption{Exponential fits to the  jump velocity distributions
for the high magnetic field (M = 5, A = 1) MHD 
256$^3$ simulation. A mean value for the number of shocks in each
direction  has been taken. The 5 times shown are t = 1,2,3,6 and 10. 
The fitted exponentials are described in the text with $\alpha = 0.66$
and $\beta = 0.62$.}   
\label{numbermagf} 
  \end{center}
\end{figure}

The jump number distribution function  for these simulation 
(Fig. \ref{numbermag})
possesses  exponential velocity distributions over a range in M$_j$. 
The displayed fit to the high field case (Fig. \ref{numbermagf}) is
\begin{equation} 
\frac{dN}{dM_j} = 10^{5.94}\,t^{\alpha}\,{\exp}(-{\beta}\,M_jt^{\alpha}).
\label{mhd-eqn}
\end{equation}
where  $\alpha \sim 0.66 \pm 0.03$ and $\beta \sim 0.62 \pm 0.03$. For the
A = 5 case, $\alpha \sim 0.57 \pm 0.03$ and $\beta \sim 1.01 \pm 0.05$.

This analysis indicates that (1) supersonic MHD turbulence is, 
{\em mathematically} at least, no different from hydrodynamic
turbulence in that the shock distribution is exponential and (2) 
the time dependencies are also similar to the hydrodynamic M = 5 case.
From Fig.\,\ref{numbermag}, we conclude (1) that the distribution of 
high speed  shocks remains unchanged and isotropic in the low-field 
case, (2) the distribution of transverse waves is somewhat faster to 
decay when a weak field is present, whereas in the strong field case 
(3) the high speed transverse waves survive significantly  longer from 
the outset  and (4) the whole spectrum of parallel waves is suppressed 
by a factor $\sim 2$.

The velocity jump distributions plotted here are a combination of 
shocks and waves. Due to the high Alfv\'en speed, the shocks are 
predominantly slow shocks in the A = 1 case. These shocks
can propagate with wave vectors in all directions except
precisely transverse to the field. Their propagation speeds are 
relatively slow since the (initially-uniform) Alfv\'en speed is 5 times 
the sound speed), and therefore their energy may be transferred into 
faster waves via the turbulence raised in the magnetic field. In any 
case, it appears from Fig.\,\ref{numbermag} that about two-thirds of 
the compressional wave/shock energy is in transverse compression modes 
for the case A = 1. The waves in this case, as measured here by regions 
of convergence along the axes, are fast magnetosonic waves (close to 
compressional Alfv\'en waves). The proportion of each can be estimated 
from the simulations by calculating the jump widths (in practice, we 
here place the extra requirement that the one-dimensional jump across 
each individual zone  exceeds 0.2\,c$_s$, in order to distinguish the 
shocks from the waves). We then  find that at M$_j$ = 1, just over half
the jumps are slow shocks (with an average resolution of $\sim 3.0$ and 3.3
zones transverse and parallel, respectively), while for   M$_j$ $> 4$, 
76\% of the 'jumps' are  actually waves  (an average of $\sim 6.0$ and 7.7 
zones in each converging region). This contrasts with the hydrodynamic
simulations where, quite uniformly, well over 90 \% of the jumps are
indeed narrow shocks, resolved only by the artificial viscosity.
These are of course  estimates which ignore the possibility  that many 
flow regions may be quite complex combinations of waves and shocks.

An explanation of why such different types of turbulence
decay in the same functional manner is offered in Sect. 6.5. 

\section{The probability distribution functions} 

A traditional aid to understanding numerically-created turbulence is
the probability distribution function (PDF) of the velocities. Here, 
we determine the one-dimensional Mass Distribution Function by calculating 
the mass per unit Mach number interval of the motion in the x-direction. 
Note all zones contribute here, whether in the shocks or not. 

\begin{figure}
  \begin{center}
    \leavevmode
    \psfig{file=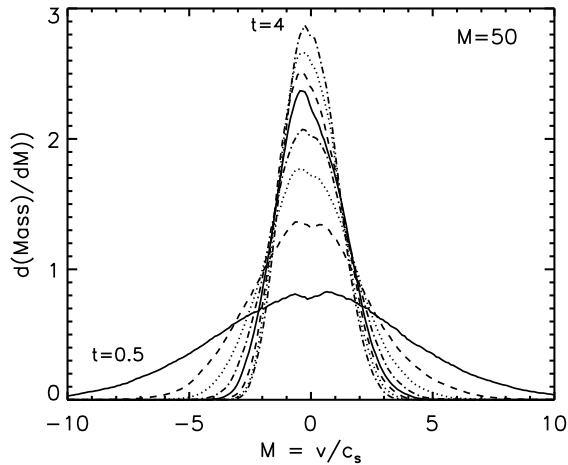,width=0.58\textwidth,angle=0}
\caption{The decay of the PDF for the M=50 simulation.
The distributions are shown for the 8 equal time steps from 0.5 to 4}   
\label{gauss}
  \end{center}
\end{figure}

\begin{figure}
  \begin{center}
    \leavevmode
    \psfig{file=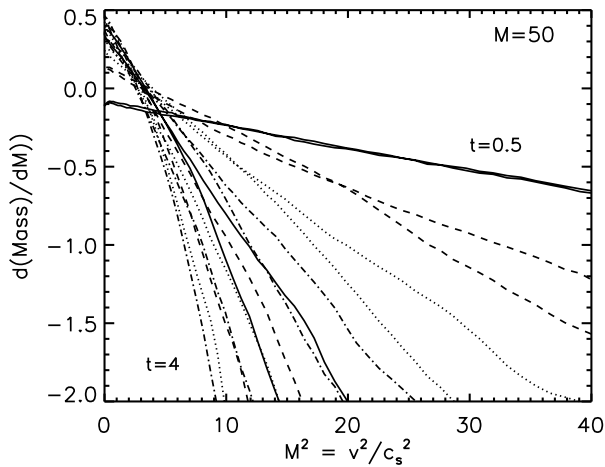,width=0.58\textwidth,angle=0}
\caption{A log-v$^2$ display of the decay of the PDF for the M=50 simulation
demonstrates the Gaussian nature of the PDFs.
The distributions for positive and negative absolute speeds
 are shown for the 8 equal time steps from 0.5 to 4, in descending order.}   
\label{gausslog}
  \end{center}
\end{figure}

\begin{figure}
  \begin{center}
    \leavevmode
    \psfig{file=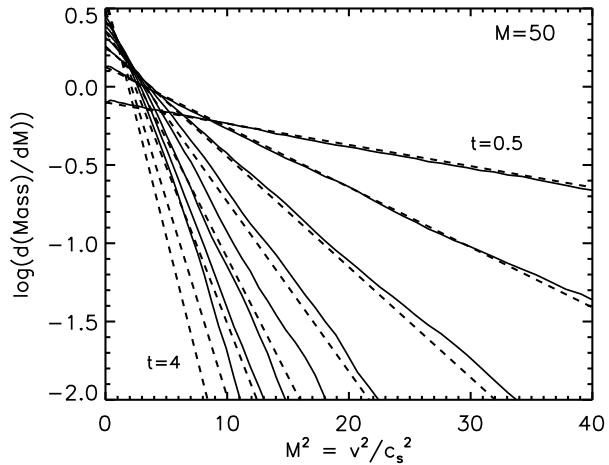,width=0.58\textwidth,angle=0}
\caption{Fits (dashed lines) to the  decay of the PDF for the M=50 simulation.
The mean distributions (full lines) are shown for the 8 equal time steps 
from 0.5 to 4. The model fits are given by Eq.\,\ref{gausseqn},
which is tailored so that the total mass in the box is conserved.}   
\label{gaussfit}
  \end{center}
\end{figure}

We recover Gaussian distributions in the velocity, as apparent 
in Fig.\,\ref{gauss}. This is clearly displayed on a Mass-log(M$^2$)
plot for the high-speed wings (Fig.\,\ref{gausslog}) where a Gaussian
would generate a linear relationship. Note
that each time step produces two lines, one for positive and one
for negative absolute velocities. At early times, the imposed
symmetry is still dominant but later on, small asymmetries
become more apparent. To estimate the time dependence we have 
taken the mean mass fraction of each pair, on defining the
initial mass density to be unity (i.e. a unit mass is initially
contained in a box of size L$^3$)
and found that a fit of the form
\begin{equation} 
\frac{d(mass)}{dM} = 1.6\,t^{0.75}\,{\exp}(-0.088\,M^2t^{1.5})
\label{gausseqn}
\end{equation}
is reasonable (Fig.\,\ref{gaussfit}). Interestingly, the 
decay in time of the PDF is not exponential. It is a faster decay
law than for the shocks. This is inherent to the nature of 
decaying turbulence: in the beginning, closely-following
shocks accelerate the fluid to high speeds. At late times, the 
fast shocks are quite evenly spread out and do not combine to 
produce high acceleration.

\section{Interpretation of shock number distribution}

\subsection{The mapping closure method}

A satisfying understanding of turbulence has been elusive. One can hope
that supersonic turbulence may possess some simplifying aspects. We
have thus devoted much time trying to interpret the above shock distributions
of decaying turbulence. In this section, we first relate the numerical 
simulations to theoretical models which have predicted exponential 
velocity gradient PDFs for other forms of turbulence. We then
interpret the evolution of the shock PDFs with an extension of these
models utilising the ballistic (ram pressure) principles behind 
hypersonic turbulence. 

We require a model based on local interaction in physical space in
order to model shock interactions. We adapt the heuristic `mapping
closure' model in the form presented by Kraichnan (1990), in which
analytical approximations were used to describe the evolution of Burgers
turbulence. First, the initial competition between the squeezing and viscous 
processes is followed. An assumed Gaussian reference field is distorted 
non-linearly in time into a dynamically evolving non-Gaussian field.
Velocity amplitude and physical space are remapped by choosing
transformations of particular forms. The PDF of a 
two-point velocity difference changes smoothly from Gaussian at very 
large separations (relating independent points) to some function 
$\xi$ at small distances. The mapping functions are then determined 
by matching the evolution equations with the dynamical equations. 
The closure is obtained by limiting the form of the distortions to
locally determined transformations.

The model for Burgers turbulence provides our inspiration since, in the
hypersonic flow simulations of isothermal gas,
thermal pressure is only significant within the  thin shock fronts.
Furthermore, individual shock structures are predominantly one-dimensional. 
Care must be exercised, however, since regions of vorticity are created behind
curved shocks which are absent in the one-dimensional Burgers turbulence.

\subsection{Background formulation}

We assume a known {\em reference field} for the initial velocity 
${\bf u_o}({\bf z})$ as a function of a {\em reference} coordinate 
system ${\bf z}$ (Gotoh \& Kraichnan 1993). The  'surrogate' velocity 
field is ${\bf u}({\bf x},t)$ and is related to ${\bf u_o}$ and the 
reference coordinates by vector and tensor mapping functions X and J:
\begin{equation}
u_i = X_i({\bf u_o},t),
\end{equation}
\begin{equation}
{\partial}z_i/{\partial}x_j = J_{ij}({\bf u_o},{\bf \xi_{o}},t)
\end{equation}
where reference velocity gradients are 
$\xi_{ij,o} = \partial{u_{i,o}}/\partial{z_j}$ and 
velocity gradients are $\xi_{ij} = \partial{u_{i}}/\partial{x_j}$.
The goal is to find the distortions described by X and J for which 
the surrogate velocity field is a valid approximation to the velocity 
field as given by the equations of motion. The above transformations,
however, constrain the allowed forms of higher order statistics and, hence,
neglect some physics which affect the long-term evolution (Gotoh \&
Kraichnan 1993).

Mapping closure also assumes that the velocity PDFs $P(u_o)$ and $P(u)$
are related to multivariate-Gaussians through prescribed forms. The 
justification is simply the statistical mechanics argument, as applied 
to particle speeds in equilibrium thermodynamics. The velocity gradient 
PDF is written as $Q({\xi},t)$ with $u$ and $\xi$ taken to be statistically 
independent. 

For convenience, we concentrate on the component of the 
velocity, $u_i$, in the $x_i$ direction. We rewrite the decay of velocity 
amplitudes in the simpler form 
\begin{equation}
u_i = X_i({\bf u_o},t) = r_i(t)\, u_{i,o}(z_i),
\end{equation}
and the velocity gradients then map through
\begin{equation}
\frac{\partial{u_{i}}}{\partial{x_i}} = \xi_{ii}  = 
                         r_{i}(t) \xi_{ii,o}J_{ii}({\bf u_o},{\bf \xi_o},t)
                         = Y_{ii}(\xi_o,t),
\label{eqnxi}
\end{equation}
so defining Y. The velocity gradient PDF, which contains the shock PDF, 
can now be written
\begin{equation}
Q(\xi_{ii}) = Q_o(\xi_{ii,o})\left[\frac{\xi_{ii}}{\xi_{ii,o}}\right]^{-1}
             \frac{N}{J_{ii}}
\label{qpdf}
\end{equation}
where $N(t)$  normalises the PDF. This is the framework in which we can 
discuss the statistical evolution of velocity gradients.

\subsection{Dynamical input}

The momentum equations being solved, with the pressure gradients  
neglected, are 
$Du_i/Dt = (1/{\rho})({\partial}/{\partial}x_j)\mu\sigma_{i,j}$  
where $Du_i/Dt = {\partial}u_i/{\partial}t + u_j{\partial}u_i/{\partial}x_j$,
$\sigma_{i,j}$ is the stress tensor and $\mu$ is the viscosity. 
Differentiation yields the velocity gradient equation:
\begin{equation}  
\frac{D\xi_{ii}}{Dt} + \xi_{ik}\xi_{ki} = 
- \frac{1}{\rho} \frac{\partial}{{\partial}x_i}
\frac{\partial}{{\partial}x_j}(\mu\sigma_{i,j})
\label{etaij}
\end{equation}
where the usual summation rule applies to j and k (but i is a chosen 
direction). 

To continue, we can derive the evolution of the functions J, by
equating $Q({\xi},t)$ as derived from these equations (yielding a
rather complex form of the  reduced Liouville equation, see Gotoh
\& Kraichnan 1993) with the Q derived from the mapping closure 
approximation (Eq.\,\ref{qpdf}). Then, however, the analysis becomes 
mathematically dense and numerical solutions are probably the best option.
 
We here revert to a simple heuristic form of mapping closure,
taking $\xi$ to be any component of $\xi_{ii}$ and the mapping 
function $J = J_{ii}$ to be determined by requiring the probability 
function $Q$ as derived from substituting for $D{\xi}/Dt$ from the reduced 
Eq.\,\ref{etaij} into the reduced Liouville equation,
\begin{equation}
\frac{\partial Q}{\partial t} + \frac{\partial }{\partial \xi}
       \left(\left[\frac{D\xi}{Dt}\right]_{c:u,\xi}Q\right) = {\xi}Q,
\end{equation}
(where $[..]_{c:u,\xi}$ denotes the ensemble mean
conditional on given values of $u$ and $\xi$)  to be equal to the $Q$ derived 
from the equivalent for mapping closure (see Gotoh \& Kraichnan 1993),
\begin{equation}
\frac{\partial Q(\xi,t)}{\partial t} + \frac{\partial }{\partial \xi}
       \left(\frac{{\partial}Y(\xi_o,t)}{{\partial}t}Q(\xi,t)\right) = {\gamma}Q(\xi,t),
\end{equation}
where $\gamma = ({\partial}/{\partial}t)\,ln(N/J)$. After some manipulation, 
this yields an equation for the evolution of J of the form (see also 
Eq.\,24 of Gotoh \& Kraichnan (1993))
\begin{equation} 
\frac{{\partial}J}{{\partial}t} = -r{\xi_o}J^2 - {\mu}k_d^2J^3 + D(J^3)
\label{jeqn}
\end{equation}
where $k_d^2 = 
{\langle}(\partial{\xi_o}/{\partial}z)^2{\rangle}/\langle{\xi_o^2}\rangle$,
angled brackets denoting the ensemble mean and $D(J^3)$ is a  function 
consisting of further non-linear derivative terms of the form $J^3$. 
Also, an integral term involving $Q$ is neglected here, which {\em a 
posteriori} limits the solutions to high jump Mach numbers
$|{\xi}| > r^2\chi_o$ ($\chi_o$ being defined below)

Note that the left hand side and the first two terms on the right 
are close to the Navier-Stokes form given 
by Kraichnan (1990), with the addition of the function $r = r_i$ which 
accounts for compressibility. These terms provide the statistics of the 
field through a non-linear transformation of the initial field with known 
statistics. The final term combines together non-linear derivatives, a step 
which permits further manipulation but with the loss of information 
concerning the constants of integration.

The evolution of $J$ begins rapidly by the  steepening of large 
velocity gradients (first term on the right), until balance with 
the viscous term (second term on the right hand side) is achieved. 
Equating these two terms yields the form of the mapping function: 
$J = -r{\xi_o}/({\mu}k_d^2)$. Substitution into Eq.\,\ref{eqnxi} 
yields $\xi = -r^2(t){\xi_o}^2/({\mu}k_d^2)$. We convert the initial 
Gaussian PDF $Q_o(\xi_o)$ into  $Q_1(\xi_1)$ by using Eq.\,\ref{qpdf} 
to yield the result that the velocity gradient PDF is transformed
into an exponential function:
\begin{equation}
    Q_1(\xi_1) = \left( \frac{\langle{\xi_o^2}\rangle}
                 {8 \pi r^2 \chi_o^2}\right)^\frac{1}{2} 
      \frac{N(t)}{|\xi_1|} \exp \left[ - \frac{ |\xi_1|}{2 r^2 \chi_o} \right]
\label{xixi}
\end{equation}
where $\chi_o = \langle \xi_o^2 \rangle/(\nu k_d^2)$. This function has 
been derived for the high gradients i.e. the shocks, consistent with the  
numerical simulations.

The exponential being set up, the gradients are then determined by the
inviscid terms. Thus, the continued evolution of $J$ is described by 
the first two terms in Eq.\,\ref{jeqn} which, on substituting the 
mapping $J = {\xi}/(r\xi_1)$ yield
\begin{equation}
\frac{\partial}{\partial t}\left( \frac{\xi/\xi_1}{r}\right) =
        \xi_1\frac{(\xi/\xi_1)^2}{r}.
\end{equation}
This has an asymptotic power law solution of the form
$\xi/\xi_1 = 1/(1+t/(t_1))$, $t_1$ being a constant. 
Hence, at large t, the  velocity
gradient associated with a fluid trajectory is inversely proportional to
time, as physically plausible for a high Mach number expanding flow. 
Substituting  $\xi_1 = (1+t/t_1)\xi$ for $\xi_1$ in Eq.\,\ref{xixi}, 
and normalising the PDF through N, yields
\begin{equation}
Q(\xi) = k \frac {\xi_o}{\xi}
            \exp \left[ - \frac{ \xi t}{\chi\xi_ot_o}   \right].
\end{equation}
for large t, where $\chi = 2\,r_o^2\langle \xi_o^2 {\rangle}/(\mu k_d^2\xi_o)$
and k is a constant.
Hence, mapping closure predicts, for zero-pressure hydrodynamic flow, 
the same fast shock decay law as uncovered in the hypersonic simulations.

The above mapping closure technique provides insight into the rapid 
build up of velocity gradients and transformation to an exponential,
as well as the evolution of the exponential term. The form of the prefactor
excludes an extension to low speeds.

\subsection{A direct physical model}

The interpretation we now present is an extension of the dynamic basis 
of the mapping closure model. First, we neglect viscosity since the 
long term evolution must be derivable from purely inviscid
theory. The critical addition to the above analysis is a result of the  
simulations: the total number of zones across which the gas is converging 
remains at approximately 30\%, independent of time. This can also be 
derived from integrating Eqs. (2) and (4) over M$_j$ which yields  
constant total shock surfaces of 1.05\,10$^6$ and  0.10\,10$^6$ zone 
surface elements, respectively (the difference being mainly the factor 
of 8 more zones in the former 256$^3$ calculation). As can be seen 
from Fig.\,\ref{fitted}), the strong shocks disappear, being replaced by  
weak shocks. We interpret this empirical conservation law as due to the fluid
being contained predominantly within the layers which drive the shocks
and the looser-defined layers which drive the weaker compressional waves.
These layers interact, {\em conserving the total shock area}. This is
expected from shock theory  since the two driving layers merge but
the leading shock waves are both transmitted or reflected.
The need for layers to drive the shocks, as opposed to shocks to sweep
up the layers, is a necessity in an isothermal flow where the
pressure behind a shock must be associated with enhanced density.
Nevertheless, a  shock will sweep up and compress pre-existing density
structures ahead of it.

The decay of a single shock is controlled by the decay of the momentum 
of the driving layer. The decay of a driving layer is here modelled as due
to the time-averaged interaction with numerous other layers. These layers 
can be represented by an 'ensemble mean' with the average density $\rho$. Thus
a shock is decelerated by the thrust of other shock layers, but its mass
is not altered. Mass is not accumulated from the oncoming shock
layer, but instead remains associated with the oncoming layer. Then, a 
layer of column density $\Sigma$ will experience a deceleration of 
$\Sigma$du/dt = -$C_d{\rho}$u$^2$ where $C_d$ is a drag coefficient of 
order unity and u is the velocity jump (i.e the relative velocity of the 
layer). Integration yields the result ut/L $\sim \Sigma$/(L$\rho$) for 
times exceeding $\Sigma$/$\rho$u$_o$ where u$_o$ is the initial layer speed.

We impose three physical conditions on the shock distribution.
\begin{itemize} 
\item At high speeds we take a standard decay law for a number of 
independently-decaying layers.  The decay rate of fast shocks is  
proportional to the number of shocks present.  In this regime, 
significant numbers of new shocks are not generated.
\item  Secondly, we shall require that the {\em total} number of shocks 
(plus compressional waves, since there is no dividing line in the numerical 
simulations) remains constant. The function which satisfies both these 
conditions clearly obeys, on integrating over all shocks with speed exceeding
$v_1$,
\begin{equation}
\frac{d\,\,}{dt} \int_{v_1}^{\infty} \frac{dN}{dv}dv =
                - \kappa(v_1) \int_{v_1}^{\infty} \frac{dN}{dv}dv
\label{condition1}
\end{equation}
with the decay rate function $\kappa(v) \rightarrow 0$ as 
$v \rightarrow 0$, to conserve
the shock number. This has solutions
\begin{equation}
\frac{dN}{dv} = a\, t\frac{d\kappa}{dv}e^{-{\kappa}t}.
\end{equation}
where a is a constant.
\item The third condition we invoke is based on the above functional 
form for individual shock deceleration  for which vt is a constant. Then,
the  number of shocks above any $v = v_o(t_o/t)$ should be conserved i.e.
\begin{equation}
\int_{v_o(t_o/t)}^{\infty} \frac{dN}{dv}dv = constant.
\end{equation}
This is similar to the result obtained for the velocity gradients
in the mapping closure analysis. Integrating Eq.\,\ref{condition1} 
with this condition then yields the jump distribution function 
\begin{equation} 
\frac{dN}{dv} = \frac{N_o}{L_o} t e^{-vt/L_o},
\end{equation}
where $N_o$ and $L_o$ are constants.
\end{itemize}
This implies that we have only two constants with which  to fit, not a single
line, but a whole family of lines! Yet, remarkably, this is quite well
achieved, as shown in Fig.\,\ref{fitted}. Note that the exponential 
time-dependence is indeed correct at high speeds, but there is a linear 
time-dependence at low speeds, where the weak shocks accumulate.

\subsection{The MHD connection}

We have shown that the decaying MHD turbulence with M\,=\,5  possesses 
similar decay properties to the  M\,=\,5 hydrodynamic case despite the 
different wave phenomena involved. Slow shocks, however, possibly 
dominate the energy dissipation in the high-field A=1 case. The power 
is dissipated within shock jumps with Mach numbers in the range 
$M_j \sim 1-2$. Alfv\'en waves are an important ingredient in the 
exponential tail of the velocity jumps. In a uniform medium, Alfv\'en 
waves do not decay even when they posess non-linear amplitudes. 
However, the Alfv\'en waves in a turbulent medium will interact 
non-linearly with other Alfv\'en waves, slow shocks, and density 
structures. Each Alfv\'en 
wave moves through a mean field of other waves. Similarly, each shock 
layer propagates through the `mean field' of other shocked layers,
the basis utilised in the above physical model for the hydrodynamic case. 
Hence, the MHD wave interactions could well lead to the decay of the velocity 
jump distribution in the same manner as shocks.

\section{Conclusions} 

Diffuse gas  under various guises is subject to supersonic turbulence. 
Large-scale numerical simulations of 3D MHD now allow us to explore
many variants. Here, we have studied an isothermal gas in which 
large-scale  Gaussian velocity perturbations are introduced and freely 
decay within a 'periodic box'. Our main aim here is to analyse the 
distribution of shocks, with the ensuing aim of determining the 
observational signatures. For this purpose, we have supplemented the 
original RMS Mach 5 simulations (Mac Low et al. 1998) with hypersonic 
Mach 50 runs. Indeed, we find that the hypersonic case leads to simple 
mathematical descriptions for the shock distribution function, from 
which the M = 5 runs deviate moderately. The Mach 50 runs obey a 
steeper energy decay law and (hence) have a faster decay of the 
spectrum of shocks. The velocity PDFs remain near-Gaussian but with 
increasing asymmetry at higher speeds and lower masses. 

It should be remarked that the velocity PDFs distribution  and shock 
distributions decay at different rates. The mass fraction at high speeds 
decreases faster than the number of shocks with similar speeds. This 
indicates that the turbulence is indeed decaying from its fully-developed 
state, and the individual shock structures interfere less with each 
other as the flow evolves. That is, the saw teeth tend to become more 
regular with time. 

Further conclusions are as follows.
\begin{itemize}
\item The magnetic field tends to slow down the spectral decay as well as the
overall energy decay. Fast shocks survive longer.
\item Transverse waves of a given strength decay faster than waves
travelling parallel to the magnetic field.
\item Apart from a small initial period, the energy is not dissipated 
by the fast shocks but by the moderate shocks with jumps in the
Mach number from upstream to downstream approximately in the range 
$M_j\,=\,1\,-\,3$, even in the M = 50 simulation. This is due to the
exponential fall in fast shock numbers, combined with the relatively 
ineffective dissipation in the weakest shocks.
\end{itemize}

Our studies show that for hydrodynamic models the ratio of $H_{\nu}$ to
$dE_{kin}/dt$ stays at about 65 percent through time, but this ratio
varies more with time and has a mean value of 30 percent for MHD
models. We conclude that short wavelength MHD waves are present, and
energy loss is distributed rather than occuring primarily in thin
layers.

In the hypersonic case, we have found simple laws for the evolution
and velocity distribution of shock speeds. The same basic exponential
spectra are  also found for the very closely related high negative 
velocity gradients of Navier-Stokes and Burgers turbulence,
in both simulations and analytical theory (e.g. Kraichnan 1990, Gotoh
\& Kraichnan 1993). We have extended their heuristic 
  mapping closure model to the  present case to reproduce our computed 
spectral forms, and thus demonstrated some of the dynamical properties which 
are inherent to supersonic turbulence. 

The relevance of studies of decaying supersonic turbulence to
star-forming clouds was questioned by Mac Low et al. (1999). The
rapid decay implies that such turbulence would be hard to catch
in action within long-lived molecular clouds. Possible sites, however, 
within which decaying turbulence should prove relevant include the regions
downstream of bow shocks, clouds suffering a recent
impact and disrupted jets. An exponential distribution of shocks such 
as we find will generate very low excitation atomic and molecular spectra 
and inefficient electron scattering, leading to steep synchrotron spectra.

In a following paper, these shock spectra will be employed to calculate
emission line spectra. On comparison with driven turbulence, we may then
begin to understand the type of turbulence we observe and what may
be producing the turbulence.

\acknowledgements

MDS benefitted greatly from the hospitality of the Max-Planck-Institut f\"ur
Astronomie.  We thank E. Zweibel for advice and discussions. Computations 
were performed at the MPG Rechenzentrum Garching. JMZ thanks the American 
Museum of Natural History for hospitality.  Partial support for this 
research was provided by the US National Science Foundation under grant 
AST-9800616.


\begin{thebibliography}{99}
\bibitem[] {}Evans C., Hawley J.F. 1988, ApJ 332, 659
\bibitem[] {}Falgarone E.,  Lis D.C., Phillips T.G. et al. 1994, ApJ 436, 728
\bibitem[] {}Franco J., Carraminana A., 1999, Interstellar Turbulence, 
               CUP, Cambridge
\bibitem[] {}Galtier S., Politano H., Pouquet A., 1997, Phys. Rev. Lett.
                   79, 2807
\bibitem[] {}Gotoh T., Kraichnan R.H. 1993, Phys. Fluids A, 5, 445
\bibitem[] {}Hawley J.F., Stone J.M. 1995, Comp. Phys. Comm. 89, 127
\bibitem[] {}Kraichnan R.H. 1990, Phys. Rev Lett. 65, 575
\bibitem[] {}Lesieur M., 1997, Turbulence in Fluids, Kluwer (Dordrecht).

\bibitem[] {}Mac Low M.-M., 1999, ApJ 524, 169 
\bibitem[] {}Mac Low M.-M., Ossenkopf V., 2000, A\&A 353, 339
\bibitem[] {}Mac Low M.-M., Burkert A., Klessen R.,
                       Smith M.D. 1998, Phys. Rev. Lett. 80, 2754.
\bibitem[] {}  Mac Low M.-M., Smith M.D., Klessen R., Burkert A., 
                      1999, Ap\&SS 246, 195 
\bibitem[] {}Padoan P., Juvela M., Bally J., Nordlund A., 1998, ApJ 504, 300
\bibitem[] {}Porter D.H., Pouquet A., Woodward P.R. 1994, Phys. Fluids 6, 2133
\bibitem[] {}Smith M.D., Eisl\"offel J., Davis C.J. 1998, MNRAS 297, 687
\bibitem[] {}Stone J.M., Norman M.L. 1992a, ApJS 80, 753
\bibitem[] {}Stone J.M., Norman M.L. 1992b, ApJS 80, 791
\bibitem[] {}Stone J.M., Ostriker E.C., Gammie C.F. 1998, ApJ 508, 99
\bibitem[] {}V\'azquez-Semadeni E. 1994, ApJ 423, 681 
\bibitem[] {}V\'azquez-Semadeni E.,  Passot T., Pouquet A. 1996, ApJ 473, 881
\bibitem[] {}von Neumann J., Richtmyer R.D. 1950, J. Appl. Phys. 21, 232 
\end{thebibliography}
\end{document}